\documentclass[a4paper,fleqn,useAMS,usenatbib]{mnras}
\usepackage[T1]{fontenc}
\usepackage{ae,aecompl}

\usepackage{amsmath}
\usepackage{amssymb}
\usepackage{graphicx}
\usepackage{url}
\usepackage{fix-cm}%remove the warning of font size

\usepackage[usenames,dvipsnames]{xcolor}
\usepackage{color}

\title[DT Void]
{DIVE in the cosmic web: voids with Delaunay Triangulation from discrete matter tracer distributions}

\author[Zhao et al.]{
\parbox{\textwidth}{
Cheng Zhao$^1$\thanks{E-mail: c-zhao15@mails.tsinghua.edu.cn},
Charling Tao$^{2,1}$,
Yu Liang$^1$,
Francisco-Shu Kitaura$^3$
\& Chia-Hsun Chuang$^3$
}
\vspace*{4pt} \\
$^{1}$Tsinghua Center for Astrophysics (THCA) \& Department of Physics, Tsinghua University, Beijing 100084, China\\
$^{2}$CPPM, Universit\'{e} Aix-Marseille, CNRS/IN2P3, Case 907, 13288 Marseille Cedex 9, France\\
$^{3}$Leibniz-Institut f\"{u}r Astrophysik Potsdam (AIP), An der Sternwarte 16, D-14482 Potsdam, Germany\\
}

\date{\today}
\pubyear{2015}

\begin{document}
\label{firstpage}
\pagerange{\pageref{firstpage}--\pageref{lastpage}}
\maketitle

\begin{abstract}
We present a novel parameter-free cosmological void finder (\textsc{dive}, Delaunay TrIangulation Void findEr) based on Delaunay Triangulation (DT), which efficiently computes the empty spheres constrained by a discrete set of tracers. We define the spheres as DT voids, and describe their properties, including an universal density profile together with an intrinsic scatter.
We apply this technique on 100 halo catalogues with volumes of 2.5\,$h^{-1}$Gpc side each, with a bias and number density similar to the BOSS CMASS Luminous Red Galaxies, performed with the \textsc{patchy} code. Our results show that there are two main species of DT voids, which can be characterised by the radius: they have different responses to halo redshift space distortions, to number density of tracers, and reside in different dark matter environments.
Based on dynamical arguments using the tidal field tensor, we demonstrate that large DT voids are hosted in expanding regions, whereas the haloes used to construct them reside in collapsing ones. Our approach is therefore able to efficiently determine the troughs of the density field from galaxy surveys, and can be used to study their clustering.
We further study the power spectra of DT voids, and find that the bias of the two populations are different, demonstrating that the small DT voids are essentially tracers of groups of haloes.
\end{abstract}


\section{Introduction}

Cosmic voids are large regions in space which contain only low luminosity (mass) galaxies (haloes/dark matter). Due to the low density nature of voids, they are less affected by nonlinear gravitational effects, and thus should be less evolved, closer to the initial gaussian field of the universe \citep[cf. e.g.][]{vande1993, Sheth2004}.
Therefore, voids are powerful tracers for cosmic structure formation and cosmological parameter constraints \citep[e.g.][]{Betancort2009, Sutter2014}, probing the nature of dark energy or alternate theories of gravity \citep[e.g.][]{Lee2009, Cai2015}, and tests of the primordial non-Gaussianities \citep[e.g.][]{Song2009}. Moreover, in our companion papers Kitaura et al. and Liang et al., we show that voids present evidence of Baryon Acoustic Oscillations (BAO) signal, which provides additional information than that of galaxies, and is used as a standard ruler for measuring the distance of the Universe.

%There are several definitions of voids, with their own specifically suitable research applications, corresponding to different void finding algorithms \citep[For a comparison of different void finders, cf.][]{Colberg2008}. There are four main types of void definition based on:
Voids can be defined in many different ways, suited for different purposes,
\begin{enumerate}
\item
under-dense regions based on the smoothed dark matter (or halo/galaxy) density field \citep[][]{Colberg2005, Shandarin2006, Platen2007, Neyrinck2008};
\item
gravitationally expanding regions based on the dynamics of the dark matter density field \citep[][]{Hahn2007, Forero2009, Hoffman2012, Cautun2013};
\item
regions free of shell crossings based on phase-space tesselations of the particle distribution \citep[][]{Abel2012, Falck2012, Shandarin2012};
\item
or empty spatial regions among discrete tracers \citep[e.g.][]{Elad1997, Aikio1998, Hoyle2002, Padilla2005, Patiri2006, Foster2009}.
\end{enumerate}

The first three classes of methods require accurate estimates of the continuous density field and are thus not optimally suited for biased discrete tracers. We will focus for this reason in this work on the last class of methods, and present a new void finding algorithm (\textsc{dive}) based on the Delaunay Triangulation \citep[DT,][]{Delaunay1934} technique, which generates meshes to discretise a spatial domain.
DT is a powerful and popular tool in finite element analysis, with a variety of applications in computational geometry, geology \citep[e.g.][]{Braun2008}, biology \citep*[e.g.][]{Medek2007}, etc.
In the past two decades, DT-based tools have been extensively exploited in astronomy \citep[e.g.][]{Bernardeau1996, Marinoni2002, Pal2006, Cardiel2011, vande2011, Berge2012, Cedres2012}.
In particular, \citet[][]{Schaap2000} developed the Delaunay Tessellation Field Estimator (DTFE) method for the reconstruction of a continuous density field from a set of discrete samples, which leads to many studies on the large-scale structures of the universe, from both theory and observation aspects \citep[cf. e.g.][]{Aragon2007, Romano2007, vande2007, Platen2011, Sousbie2011, Jennings2012}.

As a geometric tool, the DT technique is naturally useful in cosmological studies, especially for cosmic voids:
1) it is parameter-free;
2) it is able to handle arbitrary shapes of domains, even with masked regions inside, which is generally the case for observations;
3) it can be very efficient with a large number of objects. 
Indeed, the Watershed Void Finder \citep[WVF,][based on DTFE]{Platen2007} and ZOBOV \citep[][based on Voronoi Diagram, the dual graph of DT]{Neyrinck2008} void finders are both related to the DT technique.
These two methods both focus on resolving voids from the underlying density field, while our method simply takes the circumsphere of triangulation cells from a set of discrete samples (i.e., galaxy or halo catalogues).
As a direct product of the DT technique, we obtain the spherical regions, which we name ``DT voids''.

%The practical applications for clustering analysis of voids of the DIVE algorithm presented in this work (see \S \ref{sec:method}), are shown in two companion papers (Kitaura et al. and Liang et al.).{\color{cyan}[I think this is repetitive with the last sentence of the first paragraph.]}

In section \ref{sec:method} we begin with a description of our void finding algorithm, and applied it to mock catalogues in section \ref{sec:apply}.
Later, we present the properties of DT voids in section \ref{sec:property}.
Then, we study the dark matter in and around voids in section \ref{sec:environment}, and void clustering in section \ref{sec:clustering}.
Finally, we conclude in section \ref{sec:con} with a discussion on future studies.

\section{The DIVE algorithm}
\label{sec:method}

In this section we present our approach based on the Delaunay triangulation, which emerges as a natural solution to define cosmic voids from discrete distributions of objects. We dub our method \textsc{dive}: {\bf D}elaunay tr{\bf I}angulation {\bf V}oid find{\bf E}r.

%\subsection{\textsc{dive}: Delaunay trIangulation Void findEr}
%\label{sec:dive}

We consider the general case of a set $P$ of points in $d$-dimensional Euclidean space $E^d$. The Delaunay Triangulation is a partition of the convex hull of $P$ into a set $T$ of $d$-simplices (triangles in 2-D space), whose vertices are points of $P$, and the circum-hyperspheres $\mathcal{C}(T)$ do not contain any of the points in $P$ in their interiors.

In the 3-D case, DT defines all the empty spheres: $S \equiv \mathcal{C}(T)$, through the associated tetrahedra: $T$ (the corresponding 3-D simplices), with the properties, also known as the Delaunay condition:
\begin{align}
&\forall \,t \in T,\quad \mathcal{V}(t) \subseteq P\\
&\forall \,s \in S,\quad s^\circ \cap P = \varnothing ,
\end{align}
where $\mathcal{V}(t)$ denotes the vertices of $t$, and $^\circ$ is the interior operator.

Within the cosmological context, the set $P$ can be any kind of 3-D distribution of matter tracers, such as dark matter particles, haloes, galaxies, or quasars. In particular, we consider in this work the partition of haloes into tetrahedra, which circumspheres we call ``DT voids''.

Note that we do not assume that cosmological voids are spherical. In our case, DT voids do not represent individual empty regions. Instead, the distribution of DT void centres (which do not have to reside inside the associated tetrahedra) and radii associated to the corresponding circumspheres defined by tetrahedra, can be used to trace the space devoid of gravitationally collapsed objects (above certain luminosity or mass threshold).

As mentioned above, there is a deep relation between the Delaunay and the Voronoi Diagram. In fact connecting the centres of the circumspheres produces the Voronoi diagram \citep[we refer to e.g.][for applications of Voronoi Diagrams in cosmology]{Icke1987, Springel2010, Lang2015}.

The Delaunay Triangulation scheme to find all tetrahedra which connect all the $N_P$ points (matter tracers in our case) proceeds by incremental construction: the vertices of tetrahedra are added one by one, together with reconfigurations of the affected tetrahedra to ensure that the Delaunay condition is fulfilled (i.e. the requirement that the circumspheres of all tetrahedra have empty interiors.). The vertices are added in a random order to reach the efficiency of $\mathcal{O}(N_P \log{N_P})$.

In practice, we rely on the publicly available {Computational Geometry Algorithms Library}\footnote{\url{http://www.cgal.org}} \citep*[\textsc{cgal},][]{Bronnimann2015, Jamin2015} to perform Delaunay Triangulation and compute the centre and radius of the circumsphere of tetrahedra based on the halo catalogues described in the next section.

Furthermore, the \textsc{dive} algorithm has been used in Kitaura et al. and Liang et al. to validate the methodology for BAO detection from voids, and applied to observations and large sets of mock catalogues to estimate the uncertainties on such measurements.

\section{DIVE on halo catalogues}
\label{sec:apply}
To investigate the performance of the \textsc{dive} algorithm and properties of the cosmic DT voids we make use of large cosmological halo catalogues, which can be considered as proxies for Luminous Red Galaxy (LRG) catalogues, described below in \S \ref{sec:ref}, then we give a qualitative evaluation of the outcome of \textsc{dive}.

\subsection{Halo catalogues}
\label{sec:ref}

In our study, the mock halo catalogues together with the associated dark matter density field were constructed with the PerturbAtion Theory Catalogue generator of Halo and galaxY distributions \citep[the \textsc{patchy}-code,][]{Kitaura2014}.

In particular, we use the public input parameters of \textsc{patchy} \citep[][]{Kitaura2015}, which has been calibrated with the Spherical Overdensity (SO) halo catalogue (including sub-structures) of the BigMultiDark $N$-body simulations \citep[][]{Klypin2014} at redshift $z=0.56$, within the framework of the Planck $\Lambda$CDM cosmology with $\{\Omega_{\rm M} = 0.307115, \Omega_b = 0.048206, \sigma_8 = 0.8288, n_s = 0.96\}$, and the Hubble parameter $H_0 \equiv 100\,h\,\mathrm{km}\,\mathrm{s}^{-1}\mathrm{Mpc}^{-1}$ given by $h = 0.6777$. The side length of the simulation box is 2.5\,$h^{-1}$Gpc, and the grid size $960^3$ (i.e., $960^3$ dark matter particles).
Nevertheless, only the dark matter field \citep[based on augmented Lagrangian perturbation theory: ALPT,][]{Kitaura2013} on a mesh with resolution 2.6 $h^{-1}$ Mpc is taken and the haloes are populated following the explicit Eulerian nonlinear and stochastic bias prescriptions in the \textsc{patchy}-code, permitting an accurate description of the clustering obtained with the reference $N$-body simulation using $3840^3$ particles.

In fact, for the number density of haloes of around $3.5\times10^{-4}\,h^3\,\mathrm{Mpc}^{-3}$ used in this study, which is the typical density of LRG for the Baryon Oscillation Spectroscopic Survey\footnote{\url{http://www.sdss3.org/surveys/boss.php}} \citep[BOSS,][]{Eisenstein2011, Dawson2013}, it has been demonstrated that the two and three-point statistics are accurately matched \citep[cf.][]{Chuang2015, Zhao2015}.
Furthermore, these halo catalogues can be used to construct realistic mock galaxy catalogues with the Halo Abundance Matching scheme \citep[][]{Kitaura2015b, Rodriguez2015}.

We have generated 100 realisations of \textsc{patchy} mocks using the same input parameters with different seed initial conditions, to estimate the effects of cosmic variance in our analysis.
Furthermore, we convert the \textsc{patchy} mocks to redshift space with the distant observer approximation, to study the performance in observed redshift space. 

\subsection{Numerical performance}

We apply \textsc{dive} to the \textsc{patchy} halo catalogues containing about 5.5 million haloes each, and find that the Delaunay triangulation operations take on average $\sim 6$ minutes with a single CPU core.
In order to ensure periodic boundary conditions, we expand the boxes by duplicating haloes beyond the box boundaries. The expansion length is chosen to be slightly larger than the radius of the largest void in the box. We then remove voids with centres outside the box to obtain the final DT void catalogue. This full procedure takes on average $\sim$10 minutes for one realisation using a single CPU core, which makes the \textsc{dive} algorithm suitable for the analysis of large data sets. Given the low memory requirement ($\mathcal{O}(N_P)$), the \textsc{dive} algorithm can be trivially applied in parallel to different catalogues, such as the 100 considered in this study.

%\subsection{Void populations}
\subsection{Visualisations}

\begin{figure}
\centering
\includegraphics[width=.47\textwidth]{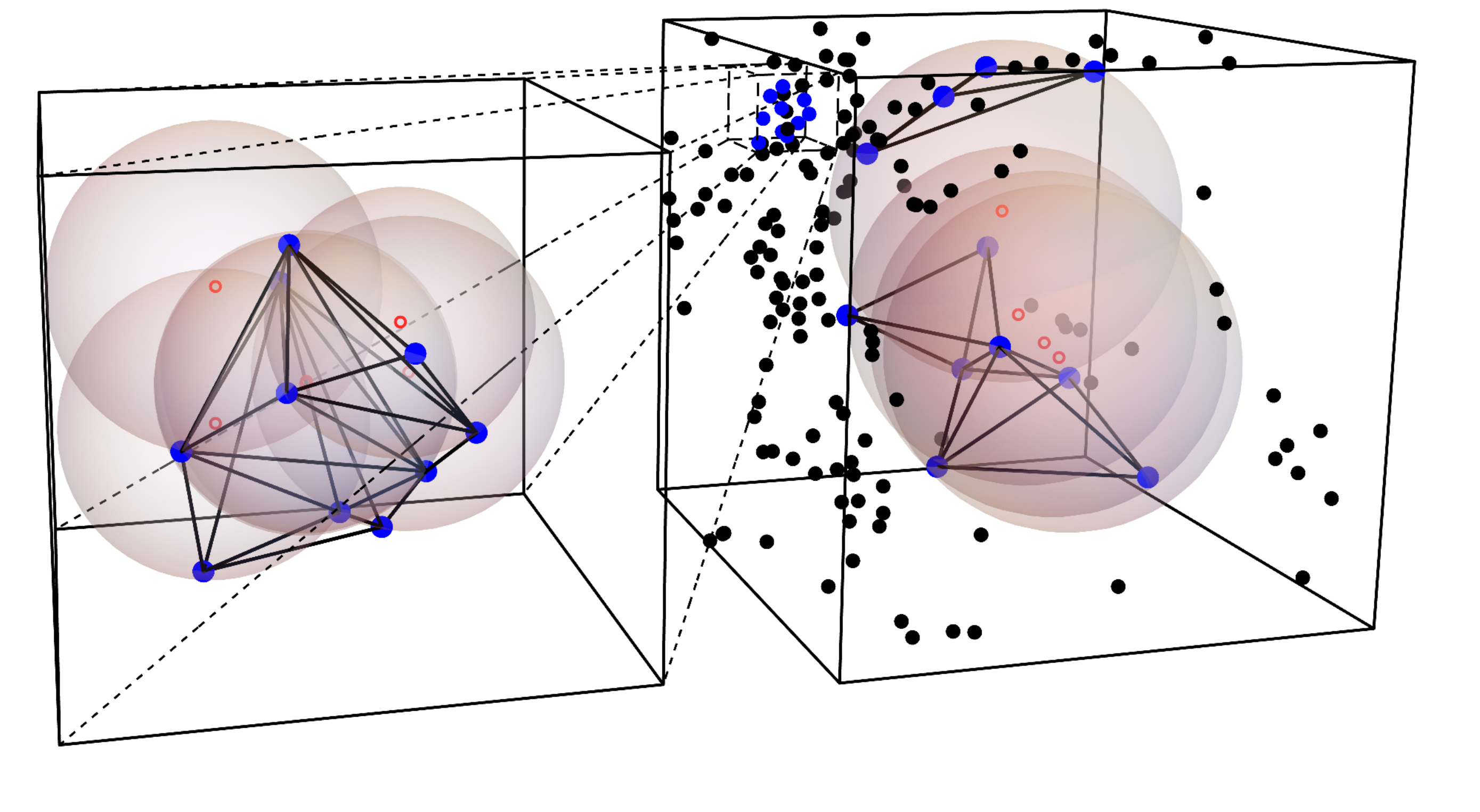}
\caption{Part of the voids resolved in a halo catalogue (black and blue points), with an extensive radius range, recognised by the \textsc{dive} technique. The red open circles denote the centres of voids, and the blue points joint by solid lines (edge of the tetrahedra) indicate the haloes associated with the voids. \textit{Left}: voids ($R_V \leq 4\,h^{-1}$Mpc) in a box with $12^3\,h^{-3}\mathrm{Mpc}^3$ volume. \textit{Right}: voids ($R_V \in [26, 27]\,h^{-1}$Mpc) in a box with $80^3\,h^{-3}\mathrm{Mpc}^3$ volume.}
\label{fig:visual_all}
\end{figure}

\begin{figure}
\centering
\includegraphics[width=.4\textwidth]{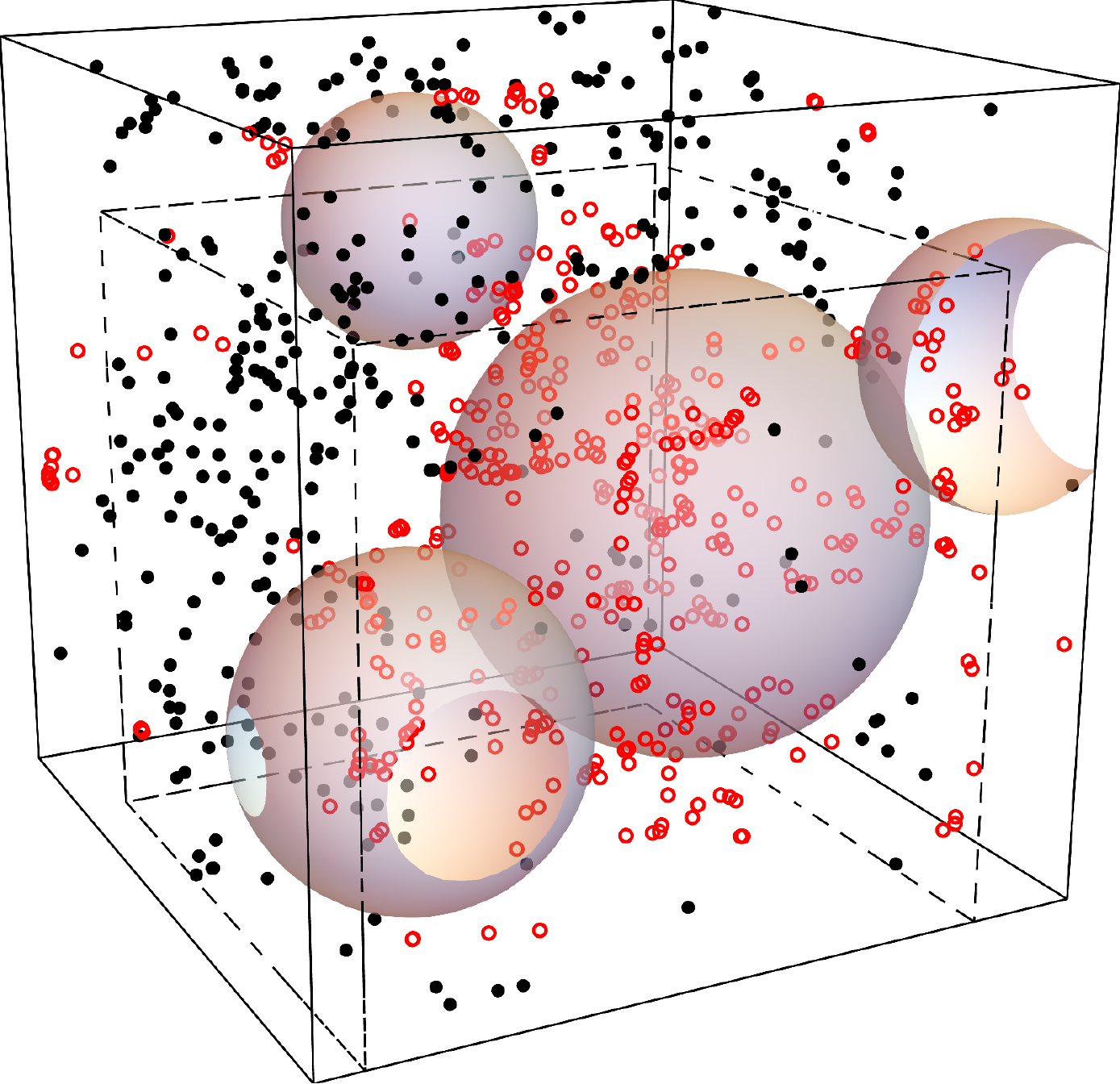}
\caption{Maximum \textit{disjoint voids} ($R_V \geq 17\,h^{-1}$Mpc) in a box with $100^3\,h^{-3}\mathrm{Mpc}^3$ volume. The red open circles indicate the centres of all the DT voids with the same radius cut. The dashed cube represents the box in the right hand side of Fig.~\ref{fig:visual_all}.}
\label{fig:visual_dj}
\end{figure}

The performance of \textsc{dive} applied to one \textsc{patchy} halo catalogue is illustrated in Fig.~\ref{fig:visual_all} and Fig.~\ref{fig:visual_dj}. 
From Fig.~\ref{fig:visual_all} we see that the centres of large voids (see the void radii of $R_V \geq 26\,h^{-1}$Mpc in Fig.~\ref{fig:visual_all}) can be distantly located to the associated tetrahedra, which indicates that they reside in low halo number density regions, while this is not the case for very small voids (see the void radii of $R_V \leq 4\,h^{-1}$Mpc in Fig.~\ref{fig:visual_all}).
Fig.~\ref{fig:visual_dj} shows that \textsc{dive} finds a large number of voids from a comparable or even smaller amount of halo tracers. This large number opens the possibility of using DT voids for clustering analyses.

\subsection{Disjoint voids}
\label{sec:disjoint}

As mentioned in the introduction, different definitions of voids can serve different purposes. \textsc{dive} can be straightforwardly applied to obtain another type of voids, the so-called maximum non-overlapping spheres \citep[][]{Patiri2006}, which have been applied to study galaxy orientation \citep[][]{Trujillo2006, Brunino2007} and proposed to constraint cosmological parameters \citep[][]{Betancort2009}. These non-overlapping spheres can be obtained from the DT voids by sorting them in decreasing radius, and sequently removing overlapping voids. 
The shaded spheres in Fig.~\ref{fig:visual_dj} show the disjoint spherical voids found by this post-processing procedure on the DT void catalogues.

From a hierarchical perspective, the \textit{disjoint voids} can be regarded as distinct (parent) nodes, and the removed overlapping DT voids as the sub-structures in concordance with the hierarchical picture of voids discussed in previous works \citep[cf.][and references therein]{Sheth2004}.
Throughout this work we refer to these two populations (or definitions) of DT voids as: \textit{all voids} and \textit{disjoint voids}.
Let us investigate more quantitatively the performance of \textsc{dive} in the coming section.

\section{Void properties}
\label{sec:property}

In this section we present a thorough analysis of the resulting void catalogues obtained from applying \textsc{dive} to the \textsc{patchy} halo catalogues which permits us to obtain insights over the DT voids and properties of voids in general.
%\subsection{void properties}
%\label{sec:population}
We start with studying the void properties such as the volume filling function \S \ref{sec:vol_func}, the number function \S \ref{sec:numdens_nmean}, and the spatial distribution of voids \S \ref{sec:spatial}.

\subsection{Volume filling fraction}
\label{sec:vol_func}

%The mass of compact objects (such as haloes) is generally used to evaluate the overdensity, similarly the volume (or filling factor) of voids can describe the emptiness of the universe. Due to the gravitational attraction force, overdensity grows with time, resulting in the expansion of voids.
Recent studies give very different Volume Filling Fractions (or cumulative volume function) of voids \citep[e.g.][]{Pan2012, Hoffman2012, Shandarin2012, Cautun2015}, varying from about 60\,\% to more than 90\,\%. However, it is commonly agreed that voids occupy most of the volume of the Universe, as gravitational attraction causes over-densities to grow with time, resulting in the expansion of voids with cosmic evolution.

We then investigate the volume filling fraction from DT voids. For \textit{disjoint voids} this is well-defined by computing the volume of the corresponding spheres. Nevertheless, it is not straightforward for \textit{all voids}, due to the high overlapping fraction, as revealed in Fig.~\ref{fig:visual_dj}. In this case we use the volume of the associated tetrahedra as a proxy to incrementally compute the volume of the corresponding void regions.

\begin{figure}
\centering
\includegraphics[width=.47\textwidth]{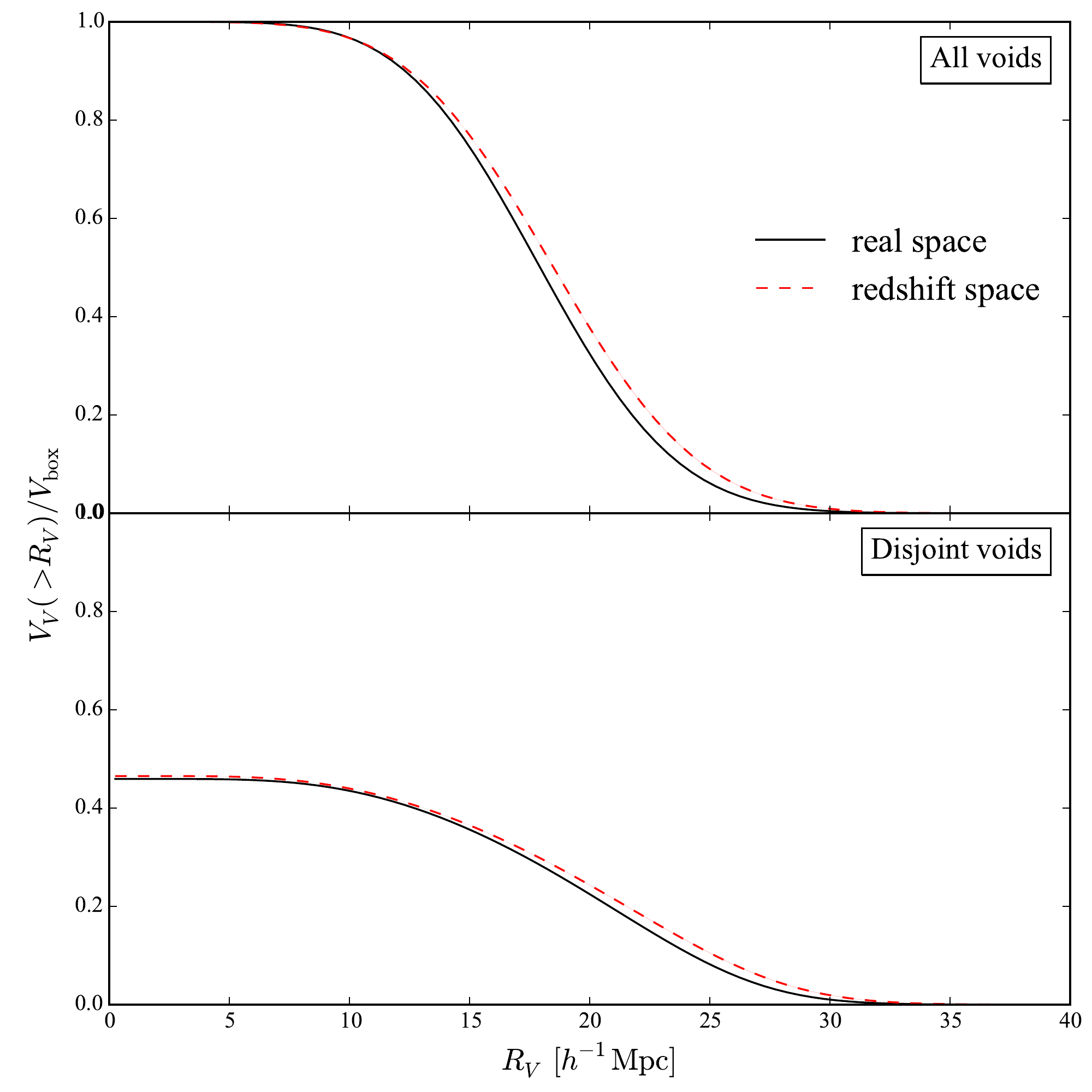}
\caption{Volume filling fraction of \textit{all voids} and \textit{disjoint voids} both in real and in redshift space obtained from the mean over 100 \textsc{patchy} boxes (the 1-$\sigma$ error bars are too tiny to be visible).}
\label{fig:volfunc}
\end{figure}

We find that the overall volume of \textit{all voids} reaches the total volume of the box, because the totality of tetrahedra fill the whole space (cf. Fig.~\ref{fig:volfunc}). The volume function of DT voids reveals the volume occupied by empty regions that can at least fit a sphere with a given radius. The volume filling fraction of \textit{disjoint voids} is much lower, covering slightly less than half of the total cubical volume considering down to the smallest voids, since cosmic empty regions are, in general, not spherical, especially at low redshifts.

Interestingly, for both species of voids, the volume filling fractions are higher in redshift space, especially at large $R_V$, because on large scales Redshift Space Distortions (RSD) squash the distribution of matter tracers along the line-of-sight, thus systematically expanding voids in an anisotropic way. RSDs on small scales, responsible for the Fingers-Of-God (FOG) effect, elongate clusters along the line of sight, hence reducing the volume of small voids. Towards small $R_V$, the volume filling fractions in real and redshift space are equal.

\subsection{Number function}
\label{sec:numdens_nmean}

The number function (or abundance) is another fundamental property of voids. It dominates the statistical errors for the analyses, and is sensitive to the cosmological parameters \citep*[cf. e.g.][]{Betancort2009, Jennings2013} and the nature of dark energy \citep*[e.g.][]{Cai2015}.

Since the DT voids are defined based on the matter tracers, the number function of voids are expected to be sensitive to the number density of haloes. To investigate this, we down-sample the 100 \textsc{patchy} mocks with halo number density of $3.5\times10^{-4}\,h^{3}\,\mathrm{Mpc}^{-3}$, and construct three more sets of \textsc{patchy} halo catalogues, with number density of $\{2, 2.5, 3\}\times10^{-4}\,h^3\,\mathrm{Mpc}^{-3}$, respectively.

\begin{figure*}
\begin{tabular}{cc}
\includegraphics[width=.47\textwidth]{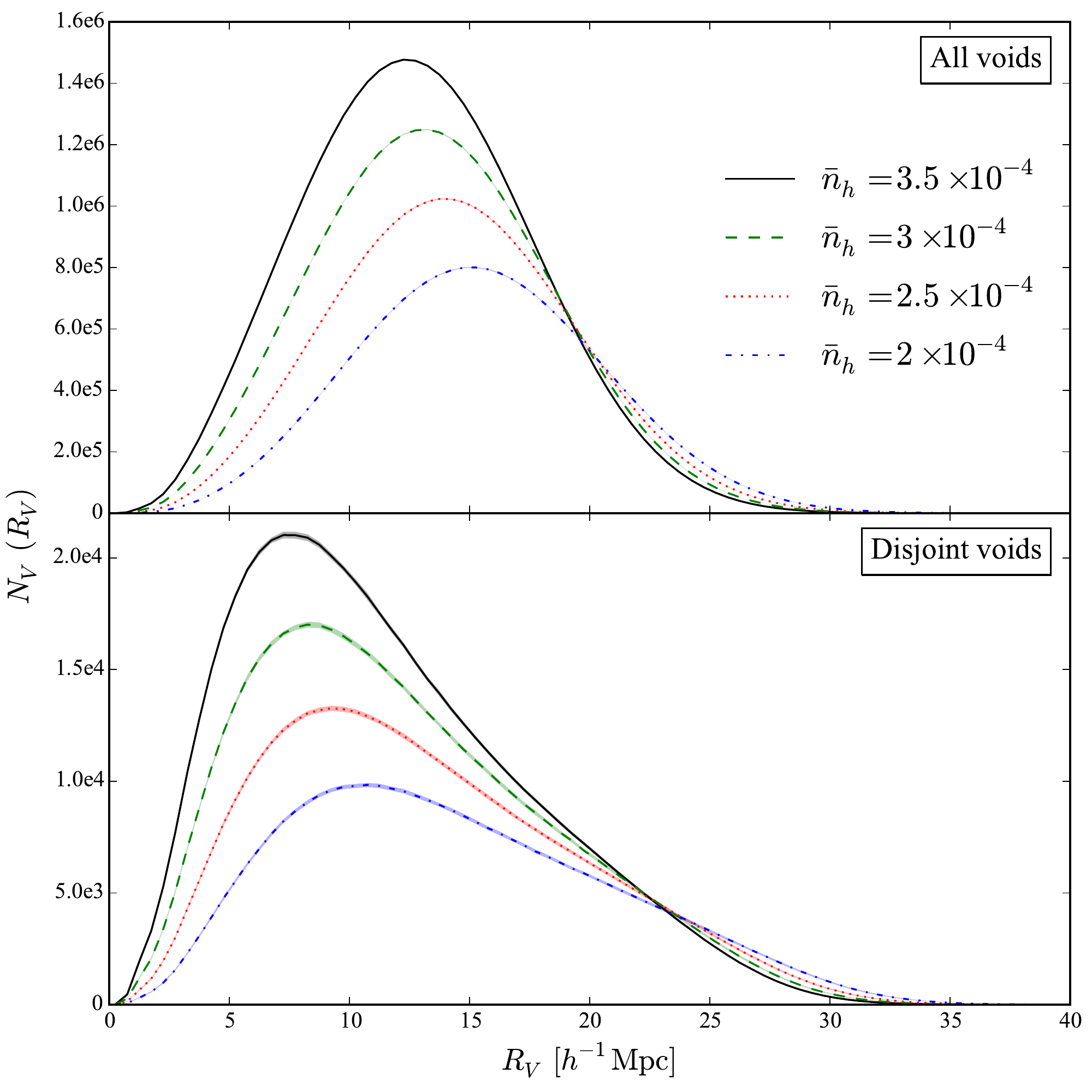}
\includegraphics[width=.47\textwidth]{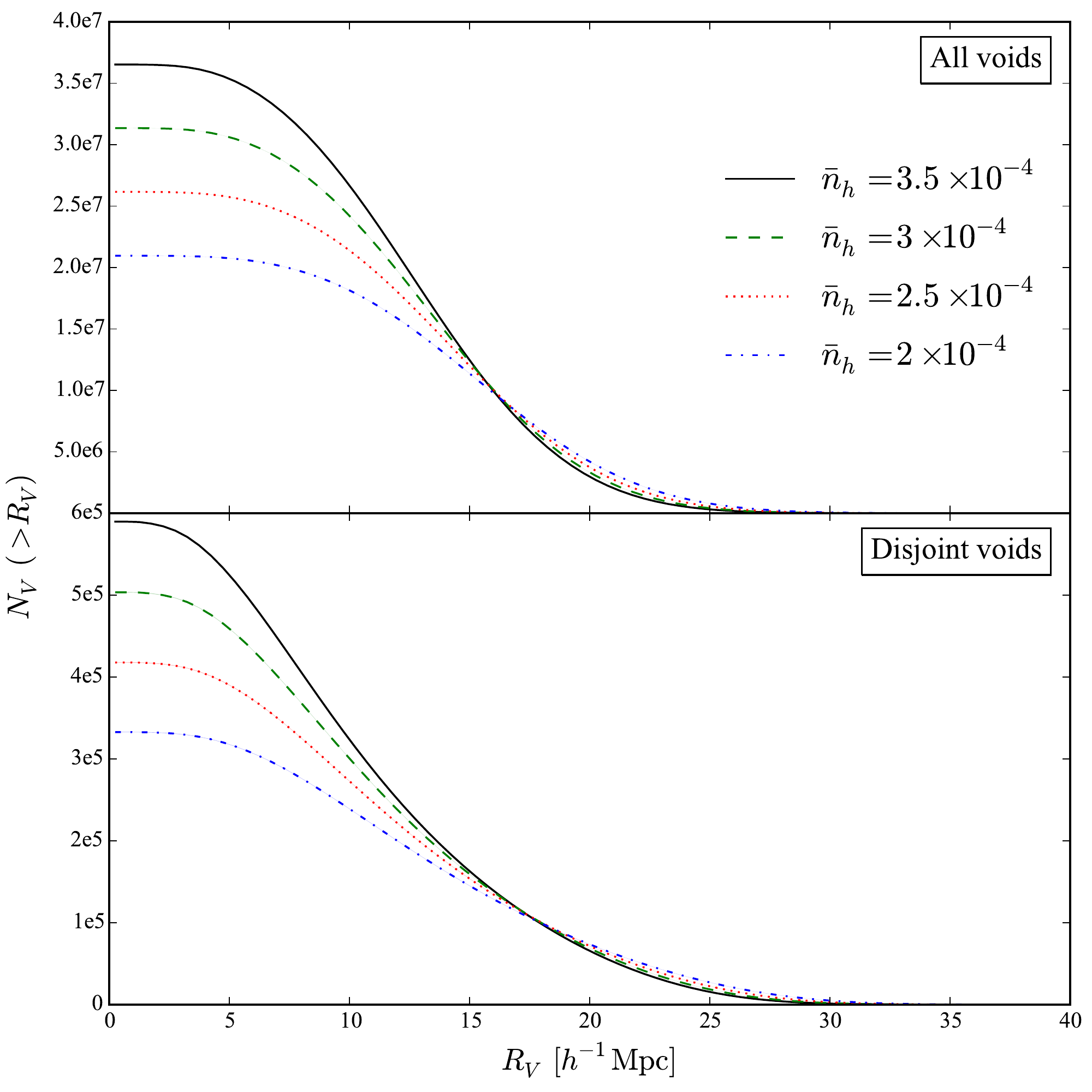}
\end{tabular}
\caption{Number function (\textit{left}) and cumulative number function (\textit{right}) of voids resolved in halo catalogues with different halo number densities $\bar{n}_h$, in unit of $h^3\,\mathrm{Mpc}^{-3}$. (The 1-$\sigma$ error bars are too tiny to be visible.)}
\label{fig:ndens_nmean}
\end{figure*}

We find that the total number of voids $N_V$, shown as a function of radius $R_V$ on the \textit{left} panel of Fig.~\ref{fig:ndens_nmean}, significantly decreases for decreasing number densities. However, the number of large voids, together with the average radius of voids, increases. This indicates that there are two populations of voids with an opposite dependency on the number density of haloes, that can be separated by the void radius: one is correlated with the population of haloes, while the other is anti-correlated with the haloes.

A visualisation of the two types of populations of voids can be found in Fig.~\ref{fig:visual_all}. The \textit{left} panel shows that the small voids are actually hosted in dense regions, while the \textit{right} panel indicates that large voids are located in empty regions. These are representing the so-called \textit{voids-in-clouds} and \textit{voids-in-voids} classification of voids, as described in \citet[][]{Sheth2004}, and are thus complementary tracers of the underlying density.

To separate the two classes of voids we consider the crossing point between the cumulative functions for varying halo number densities, as shown on the \textit{right} panel of Fig.~\ref{fig:ndens_nmean}. From this we find that the transition from one class to the other occurs for void radii of about $R_V \geq 16$ and 17.5$\,h^{-1}$Mpc for \textit{all voids} and \textit{disjoint voids}, respectively, which is not insensitive to the halo number densities.
Interestingly, the transition radius for \textit{all voids} is precisely the radius cut found to provide the optimal signal-to-noise ratio for the BAO signal from DT voids. Adding smaller radius contaminates the sample with \textit{voids-in-clouds} with the opposite BAO signature (cf. Liang et al.).

\begin{figure}
\centering
\includegraphics[width=.47\textwidth]{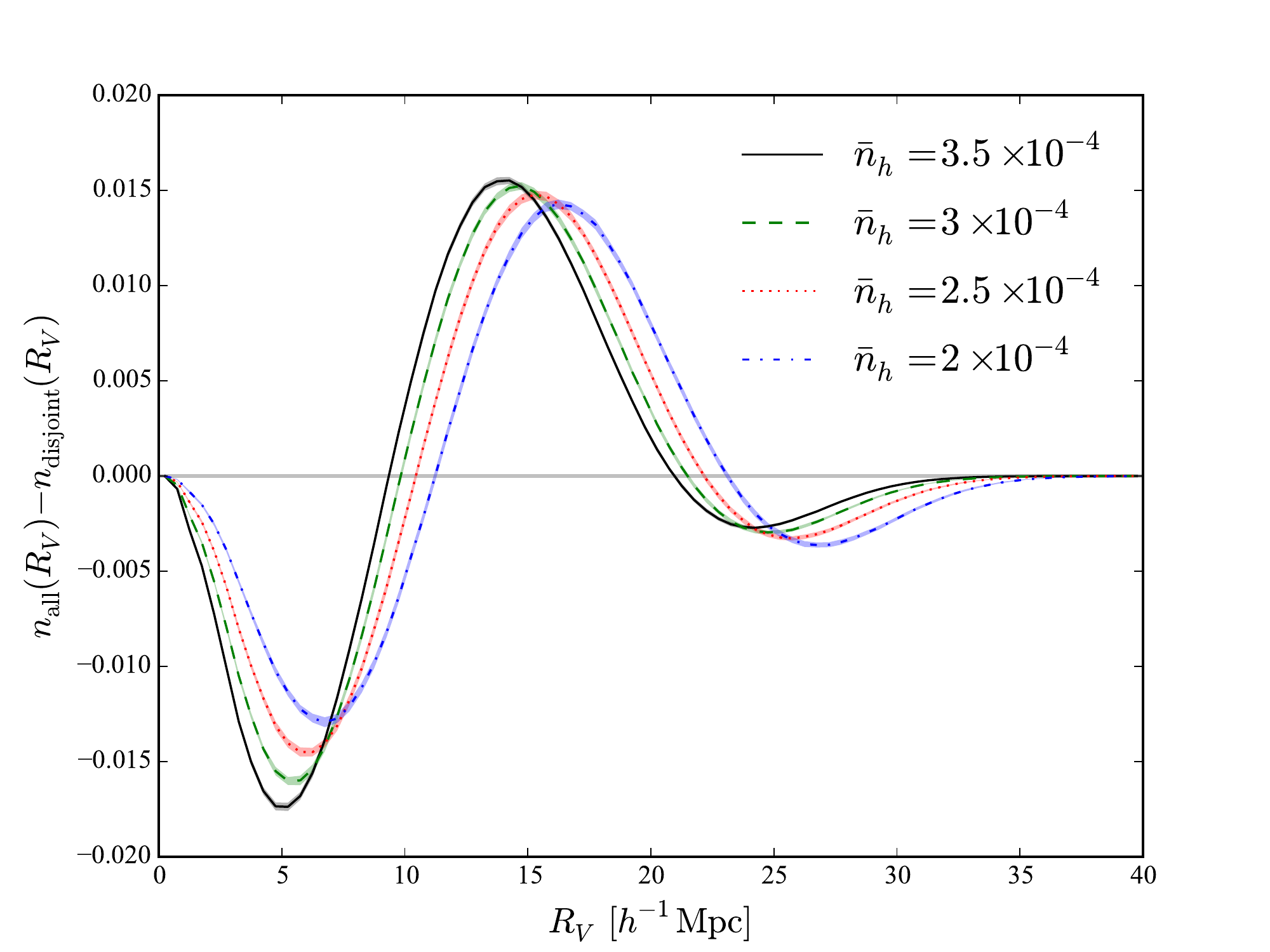}
\caption{Difference of the number fraction between \textit{all voids} and \textit{disjoint voids} resolved in halo catalogues with different halo number densities $\bar{n}_h$ in units of $h^3\,\mathrm{Mpc}^{-3}$ represented by the corresponding lines. The shaded regions show the 1-$\sigma$ errors obtained from 100 mocks with the corresponding colour-code.}
\label{fig:ndens_diff}
\end{figure}

The disjoint population has a larger fraction of small voids in general, while the critical radius for the two populations is slightly higher. The different number fraction between \textit{all voids} and \textit{disjoint voids}, shown in Fig.~\ref{fig:ndens_diff}, indicates that voids with medium radius ($R_V \in [10, 20]\,h^{-1}$Mpc) have high overlapping fractions in contrast to small voids ($R_V \leq 10\,h^{-1}$Mpc).

\begin{figure}
\centering
\includegraphics[width=.47\textwidth]{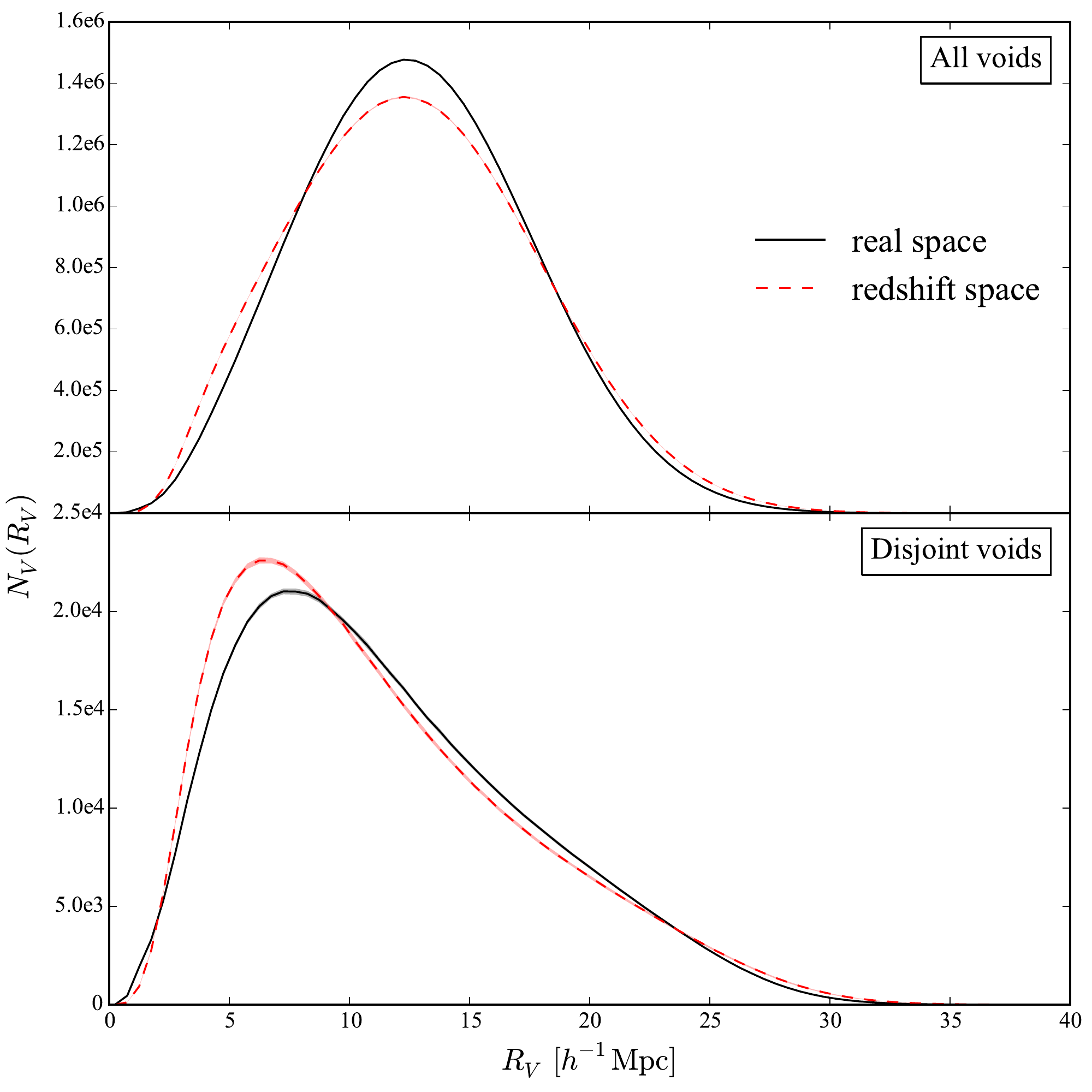}
\caption{Number function of voids in real and redshift spaces. (The 1-$\sigma$ error bars are too tiny to be visible.)}
\label{fig:ndens_z}
\end{figure}

The effect of RSDs in the halo distribution on the void number function is related to that for void volumes, and \textsc{dive} finds more voids with large radii (cf. Fig.~\ref{fig:ndens_z}). There is also a decrease in the number of voids with intermediate radii, presumably due to nonlinear RSDs (FOGs).
For smaller voids with $R_V \sim 5\,h^{-1}$Mpc, the number of voids increases again. This seems to be another effect of the \textit{voids-in-clouds}. However, we should be cautious at this point, as the details of RSDs at small scales could have some systematic deviations from $N$-body simulation, since the \textsc{patchy} realisations rely on perturbation theory, and the \textsc{patchy} dark matter fields we rely have a resolution of $2.6\,h^{-1}$Mpc.

\begin{figure}
\centering
\includegraphics[width=.47\textwidth]{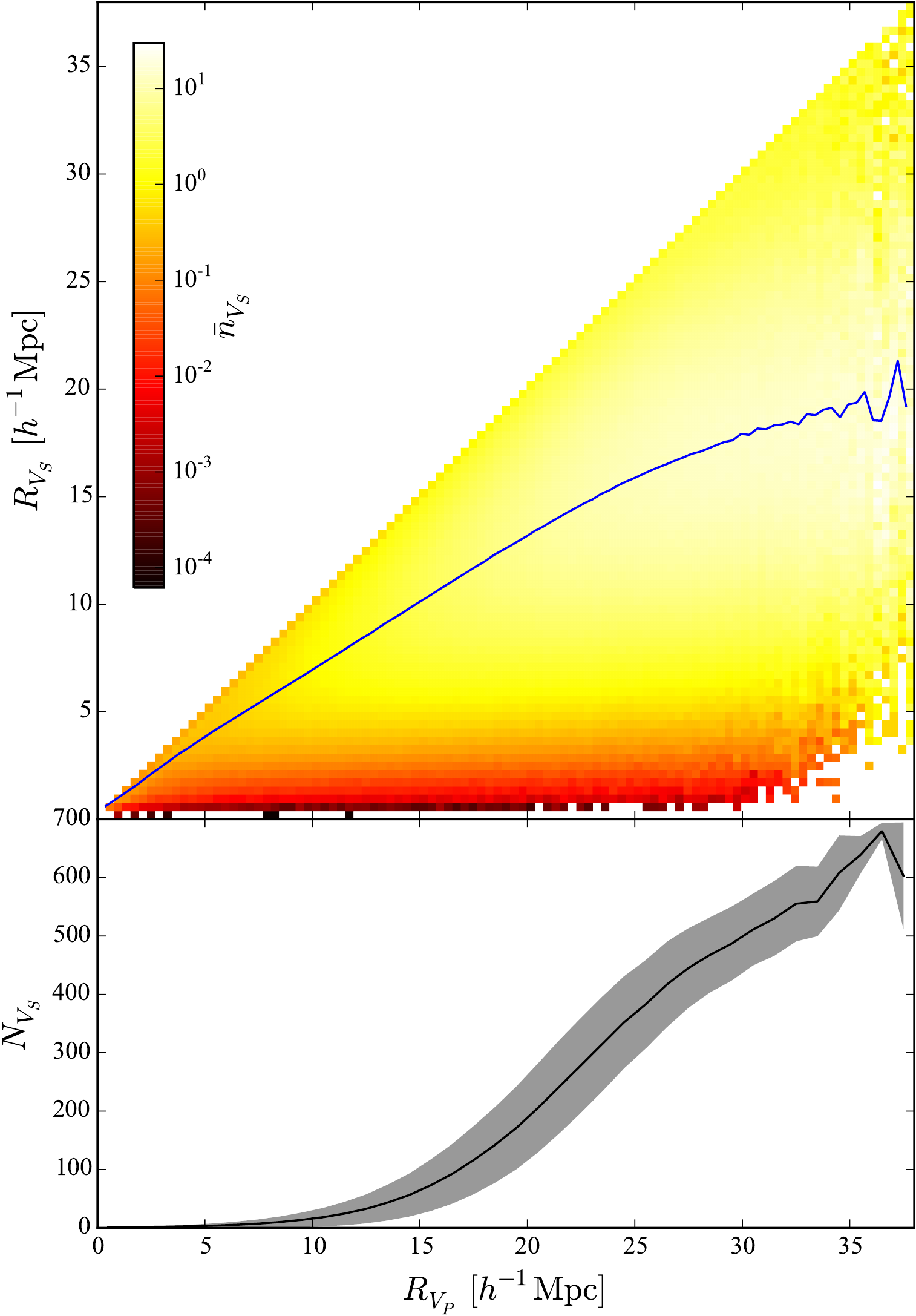}
\caption{Radius and number distribution of sub-voids for different radius of the associated parent (distinct) voids. \textit{Top}: for given radius of parent voids ($V_P$, or \textit{disjoint voids}), the mean number fraction of sub-voids ($V_S$, or overlapping DT voids) of different radii. The blue curve represents the mean radius of sub-voids. \textit{Bottom}: Number of sub-voids for different radii of parent voids. The shaded area shows the 1-$\sigma$ errors obtained from all the parent voids in one realisation of \textsc{patchy} mocks.}
\label{fig:subvoid}
\end{figure}

To further explore the characteristics of the distribution of overlapping voids, we interpret the \textit{disjoint voids} as parent voids ($V_P$), and overlapping voids as sub-voids ($V_S$). With these definitions we find that larger parent voids tend to have larger and more sub-voids (cf. Fig.~\ref{fig:subvoid}). 
By changing the void definition from \textit{disjoint voids} to \textit{all voids} with large radii, we obtain more large voids in large empty regions, giving more weights to these regions.

\subsection{Spatial distribution of matter tracers and voids}
\label{sec:spatial}

%Apart from describing the population of voids statistically, it is meaningful to see how the voids are distributed in space, as in \citet[][]{Hamaus2014b}. The large-scale distribution of structures in the universe is known as the cosmic web, in which voids are a main component. The morphology of the cosmic web is a crucial nature of the universe, and has been extensively studied \citep[e.g.][]{Platen2011, Sousbie2011, vande2011, Shandarin2012, Nuza2014}.

\begin{figure*}
\centering
\includegraphics[width=.8\textwidth]{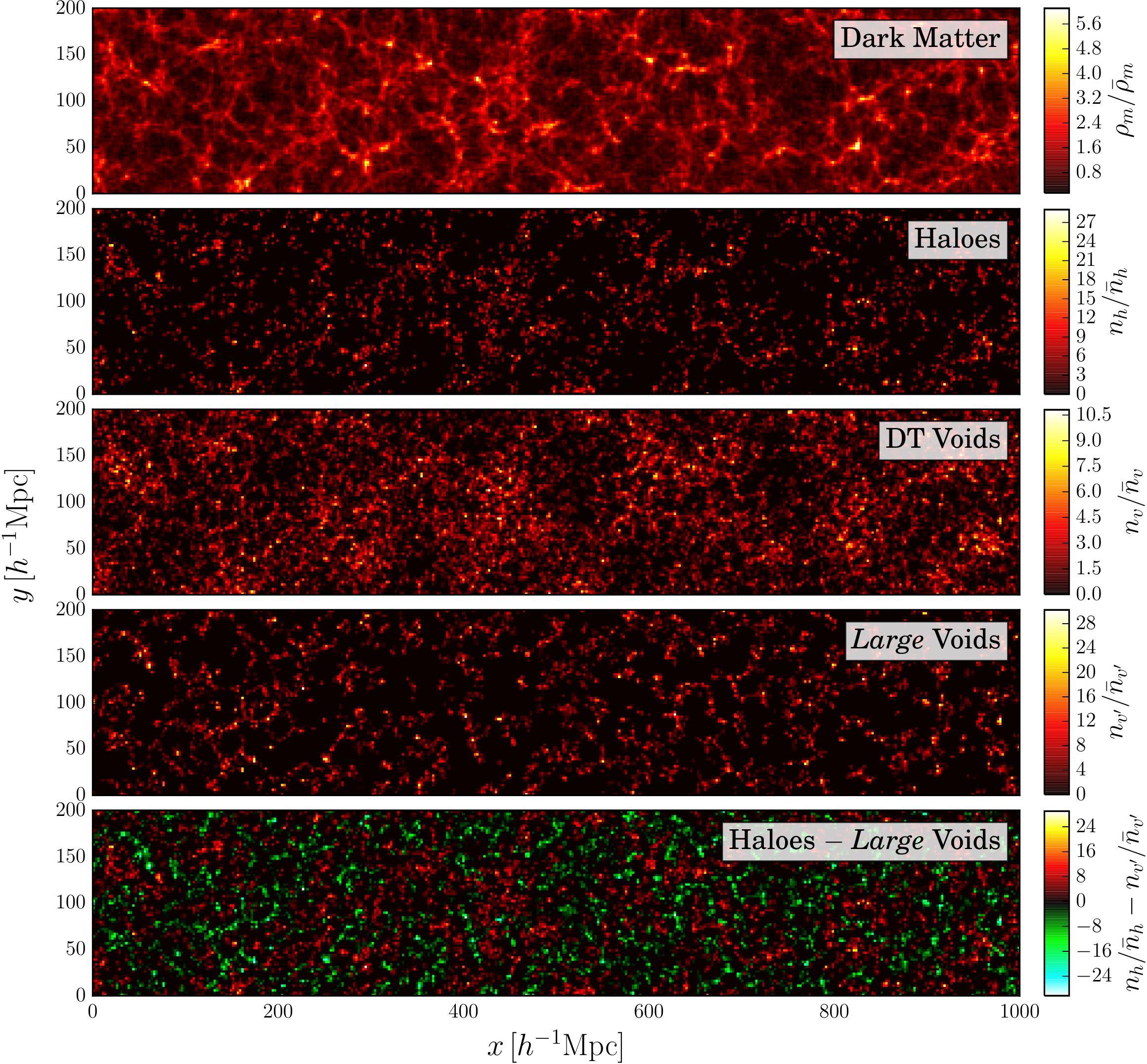}
\caption{Projected density field for different tracers in a $1000\times500\times50\,h^{-3}\mathrm{Mpc}^3$ box extracted from one \textsc{patchy} realisation. \textit{Large} Voids denote DT voids with $R_V \geq 16\,h^{-1}$Mpc. Haloes $-$ \textit{Large} Voids indicate the combination of normalised number density of haloes and \textit{large} voids, where the number density of voids is denoted by negative numbers.}
\label{fig:densfield}
\end{figure*}

In addition to the statistical description of the population of voids, we investigate the spatial distribution of voids.
Fig.~\ref{fig:densfield} \citep[inspired by][]{Hamaus2014b} presents the projected density field for different tracers extracted from a thin slice of one \textsc{patchy} realisation, where the normalised densities are shown: $\rho / \bar{\rho}$ or $n / \bar{n}$, where $\rho$ and $n$ denote the density and number density, respectively, and $\bar{\rho}$ and $\bar{n}$, indicate the average values across the whole volume.

The halo distribution follows the over-densities in the dark matter distribution, so does
the number density field of DT voids. This indicates that the \textit{voids-in-clouds} dominate the number density distribution, and is consistent with the number function shown in Fig.~\ref{fig:ndens_nmean}.

The \textit{voids-in-voids} population of DT voids (\textit{large} voids, with $R_V \geq 16\,h^{-1}$Mpc) is shown combined with the number density field of haloes with opposite signs at the \textit{bottom} panel in Fig.~\ref{fig:densfield}. Haloes and \textit{large} voids are clearly spatially separated, and together, fill the different regions of the cosmic web, forming a homogenous map. It indicates that the centres of \textit{large} DT voids indeed trace the empty regions of the underlying density distribution, while small voids follow the distribution of dense regions.
This claim needs however further quantitative verification, which we present below.

\section{Dark matter density field around voids}
\label{sec:environment}

We present in a more quantitative way the environment in which the different classes of DT voids reside.

\subsection{Local dark matter density}
\label{sec:local_dens}

The most direct cosmological environment information of the voids is given by the local dark matter density near the void centres. In order to explore the local environment of DT voids, we take the dark matter density field from \textsc{patchy} mocks, with the mesh size of about $2.6\,h^{-1}$Mpc ($960^3$ cells). The local dark matter density of each void is represented by the density of the closest cell.

\begin{figure}
\centering
\includegraphics[width=.47\textwidth]{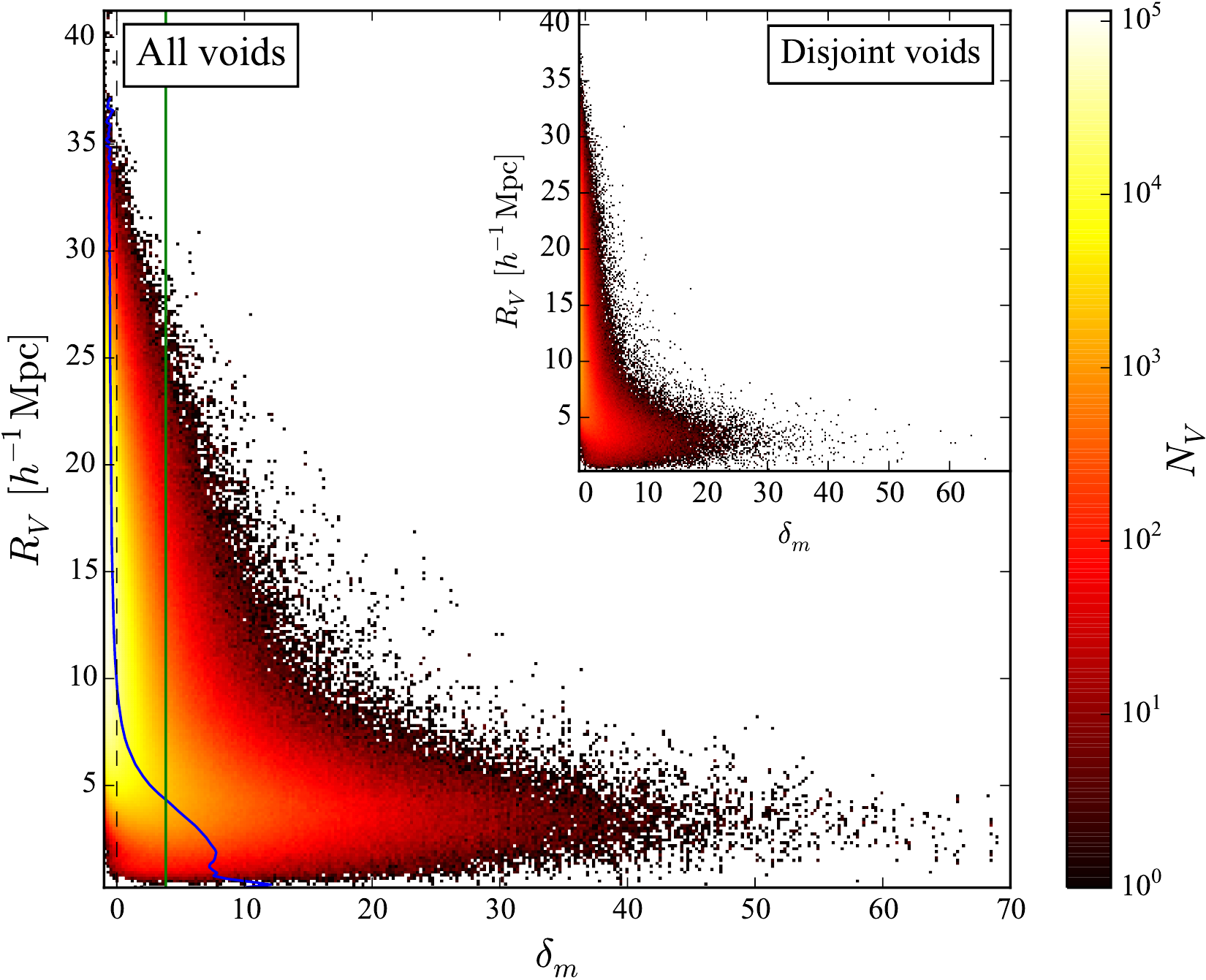}
\caption{Number of DT voids with different radius ($R_V$) and local dark matter density contrast ($\delta_m \equiv \rho_m / \bar{\rho}_m - 1$). The blue curve stands for the mean DT voids density contrast as a function of radius, while the green line stands for the mean halo density contrast.}
\label{fig:radens}
\end{figure}

The distribution of DT voids as a function of radius and local dark matter density (cf. Fig.~\ref{fig:radens}) confirms that there are two void populations, one with large radii which reside in low density regions, and the other with small radii spread over the full density range.

The dispersion of the local dark matter density steadily decreases for increasing DT void radii. The majority of these voids are hosted in low density regions, and are hence tracing the empty regions with very low density.
In contrast, the dispersion of the local density rapidly grows towards small DT void radii, reaching the typical density range for haloes. This indicates that the majority of small voids with $R_V < 5\,h^{-1}$Mpc are located in over-dense regions, and have a large bias. In particular, DT voids with $R_V \sim 4\,h^{-1}\mathrm{Mpc}$ have the same mean density as the haloes used to construct them.

\subsection{Density profiles}
\label{sec:dens_pro}

The density profile of voids is a potentially powerful signature for constraining cosmological parameters and theories of gravity \citep[e.g.][]{Cai2015}, although in practice it might be difficult to get precise measurements from observations.
The universality of void density profiles has been discussed with many different methods and data \citep[e.g.][]{Colberg2005, Ricciardelli2013, Hamaus2014a, Nadathur2015}.
However, little attention has been paid to the intrinsic scatter of the density profiles. In general, void finders based on the reconstructed dark matter density field (e.g. ZOBOV and WVF) have relatively low errors \citep[e.g.][]{Hamaus2014a}, while the fluctuations of the density profile from other void finders can be considerably large \citep[e.g.][]{Ricciardelli2013}.

\begin{figure*}
\begin{tabular}{cc}
\includegraphics[width=.47\textwidth]{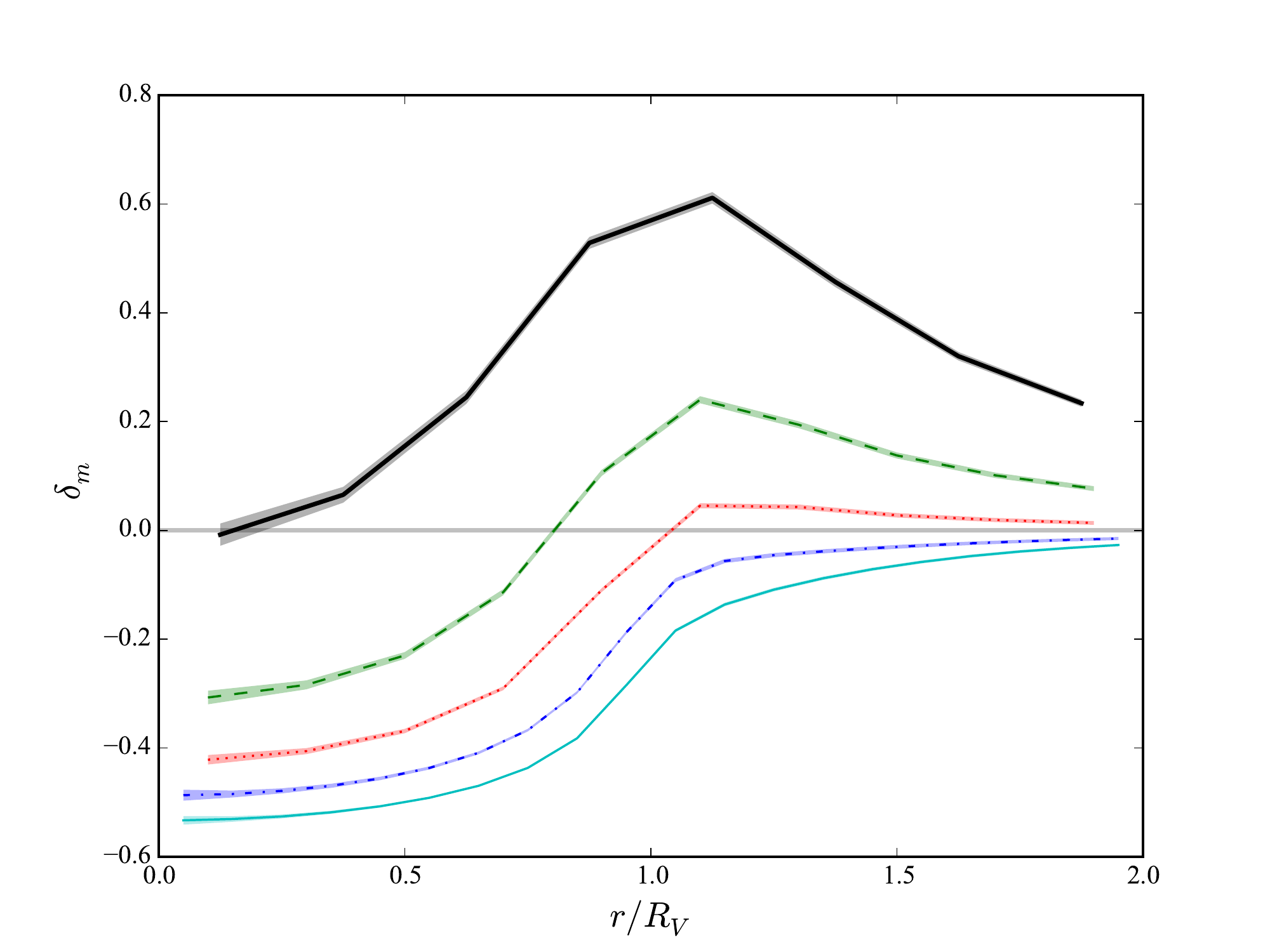}
\includegraphics[width=.47\textwidth]{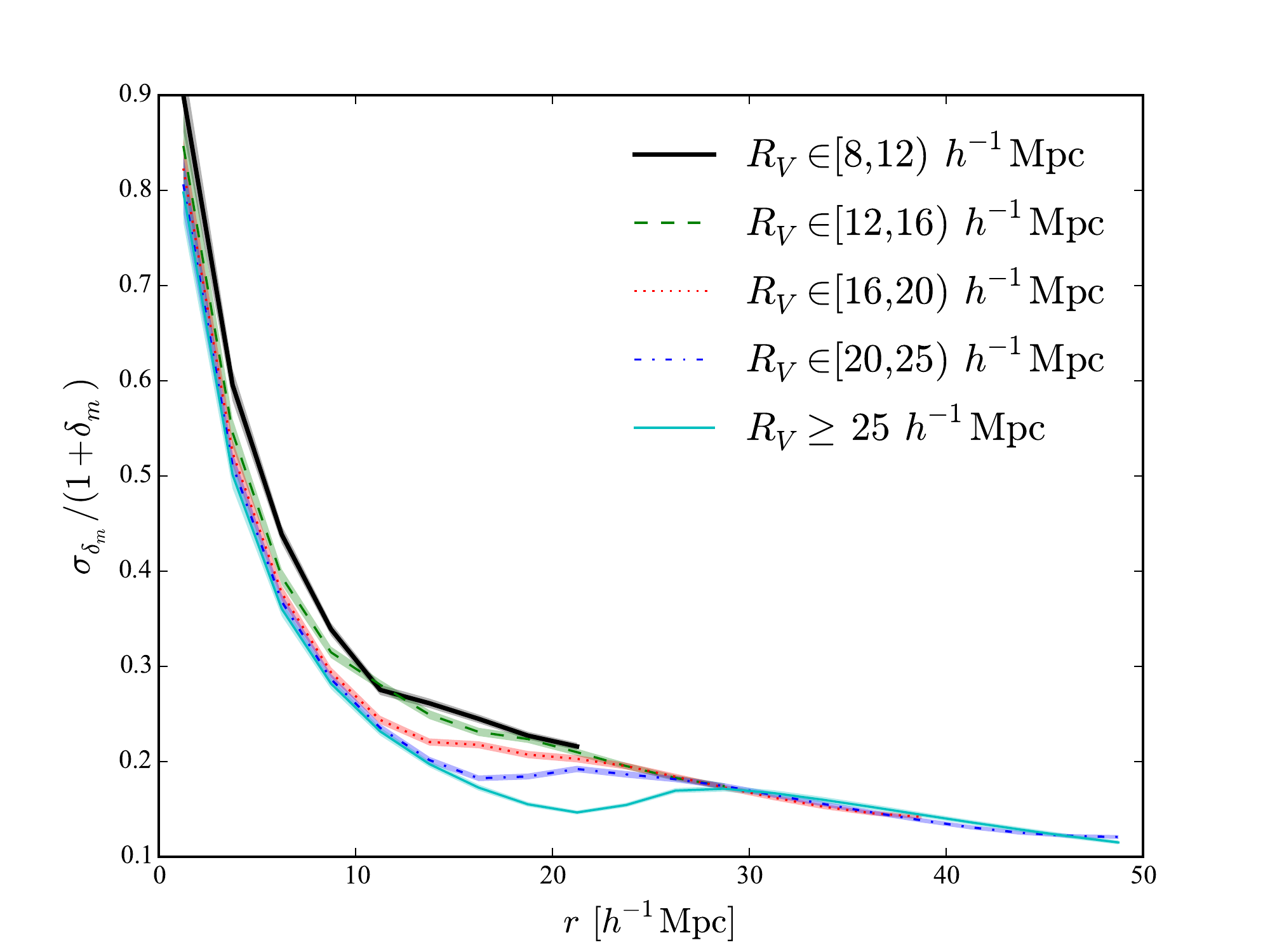}
\end{tabular}
\caption{Mean density profile (\textit{left}) and the standard deviation of the density profile (\textit{right}) of DT voids, where $r$ is the distance to the centre of the void. The shaded regions show the 1-$\sigma$ errors obtained with sub-samples.}
\label{fig:dens_pro}
\end{figure*}

Since the DT voids number density is high (cf. \S~\ref{sec:numdens_nmean}), it permits us to have reliable analyses for both the density profile and its fluctuations.
Due to the limitation of the mesh resolution ($\sim2.6\,h^{-1}$Mpc) and the accuracy of the \textsc{patchy} dark matter density field, we select only relatively large DT voids and divide them into 5 classes according to the radius (cf. Fig.~\ref{fig:dens_pro}) for the density profile studies.
Each class of voids, is further randomly partitioned into sub-samples that contain 2,000 voids. With one \textsc{patchy} realisation, the class with $R_V \geq 25\,h^{-1}$Mpc contains $\sim$140 sub-samples, while other classes with smaller radii have thousands of sub-samples. We then stack the voids in the sub-samples and compute the radial dark matter up to twice the void radius to obtain the density profile of DT voids.

We find that the shape of the DT void density profiles are in general close to the universal density profile presented in previous literature, however, our results show a stronger dependence on the radius (cf. Fig.~\ref{fig:dens_pro}). We further find that small voids, representing \textit{voids-in-clouds}, can show positive density contrasts even near their centres. 
The profiles show a steep wall near the boundary of the voids, crossing zero over-density, and then, for larger radii, the density declines steadily towards the background value.
For very large voids, we observe a similar rise in density. However, there is no zero over-density crossing close to the boundary.
These universal characteristics of void density profiles give a robust quantitative support to the initial claim that the centres of DT voids trace the emptiness of the universe. Furthermore, the volume covered by the distribution of DT voids is close to the voids defined by other sophisticated methods, based e.g. on the watershed approach.

Furthermore, the fluctuations of the density profiles normalised by the dark matter density ($\sigma_{\delta_m} / (1+\delta_m)$) also show an interesting universality, especially at small scales ($R_V \leq 10\,h^{-1}$Mpc).
By combining the sub-samples, we find an intrinsic scatter of the density profile, which does not decrease with an increasing number of voids. The origin of this fluctuation is probably the background fluctuation level of the density field in under-dense regions.
The \textit{right} panel of Fig.~\ref{fig:dens_pro} indicates an invariant behaviour of this intrinsic fluctuation. However, at scales close to the radius of the voids, such invariance is broken.

\subsection{Cosmic web environment}
\label{sec:cosweb_env}

The cosmic web is important to characterise the large scale structure and can be used to understand structure formation or test $\Lambda$CDM \citep[e.g.][]{Platen2011, Sousbie2011, vande2011, Shandarin2012, Nuza2014}.
So far we studied the relation between DT voids and the underlying dark matter density in a static perspective.
We now use the dynamical cosmic web classification method introduced by \citet[][]{Hahn2007} and \citet[][]{Forero2009}, in order to clarify the dynamical tidal field environment of DT voids, in which the cosmic web structures are classified into four types: knots, filaments, sheets, and voids.
%However, we are missing the information of the cosmic web structures, and the dynamics of the density field. In this case, we employ the dynamical cosmic web classification method introduced by \citet[][]{Hahn2007} and \citet[][]{Forero2009} (where the cosmic web structures are classified into four types: knots, filaments, sheets, and voids), aiming at clarify the dynamical tidal field environment of DT voids.
In brief, this method computes the tidal field tensor ($T_{i j}$) according to the density field ($\delta_m$), i.e.,
\begin{equation}
T_{i j} = \frac{\partial^2 \phi}{\partial x_i \partial x_j} ,
\end{equation}
where $\phi$ is the gravitational potential, and is obtained through the Poisson equation:
\begin{equation}
\nabla^2 \phi = \delta_m .
\end{equation}
Then we compare the three eigenvalues of $T_{i j}$ with a given threshold ($\lambda_{\rm th}$). If all the three eigenvalues are above (below) $\lambda_{\rm th}$, then the local structure is collapsing (expanding) on all directions, forming a knot (void). When one (two) eigenvalue is above $\lambda_{\rm th}$, we have filament-like (sheet-like) structures.

\begin{figure}
\centering
\includegraphics[width=.47\textwidth]{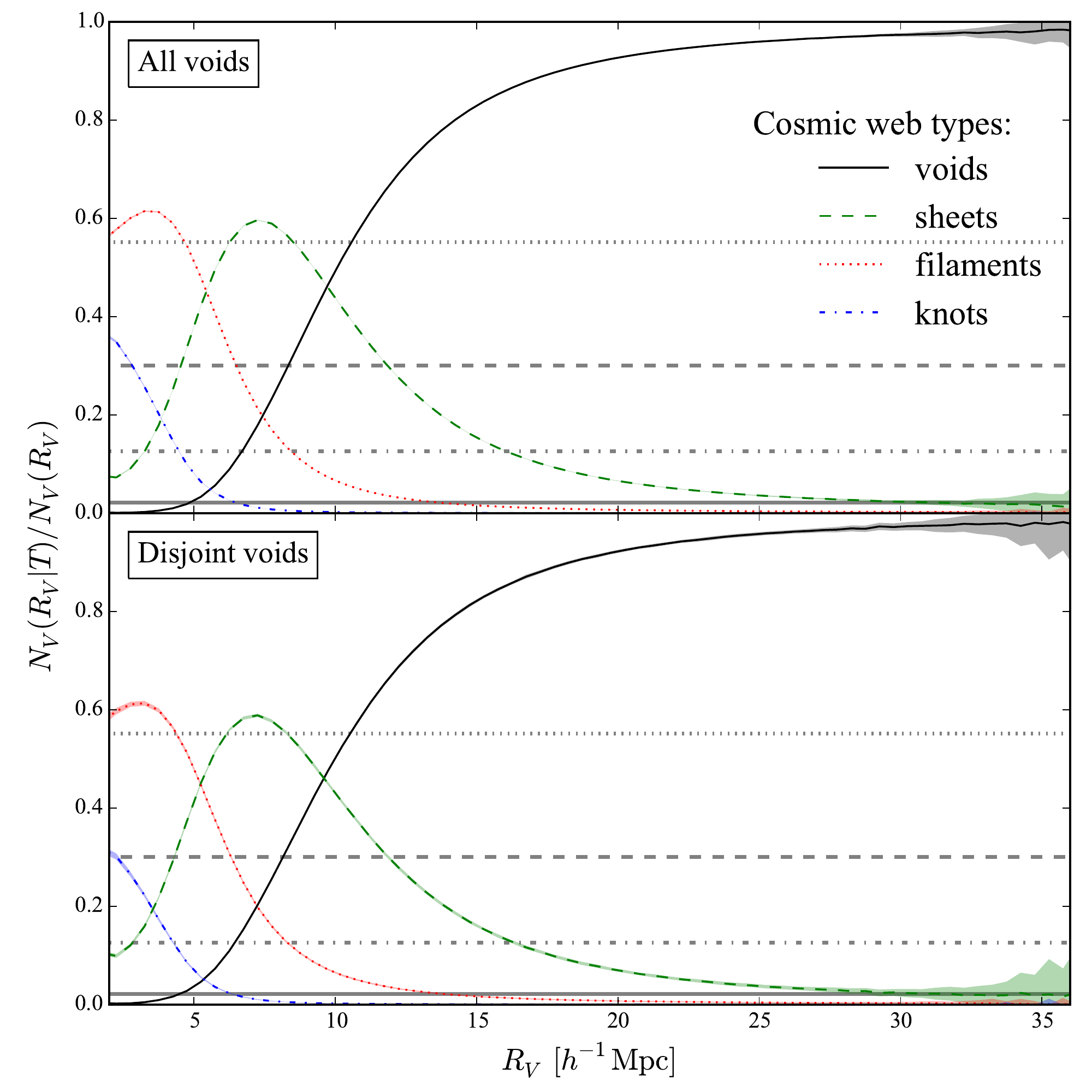}
\caption{Fraction of DT void centres for different cosmic web types ($T$) defined in a dynamical way. The thick grey lines denote the fraction of haloes in different cosmic web types. The shaded regions show the 1-$\sigma$ errors obtained from 100 mocks.}
\label{fig:cosweb}
\end{figure}

In particular, we choose the $\lambda_{\rm th}$ parameter of $0.5$ in this classification method, such that the volume filling fraction of the four cosmic web types (knots, filaments, sheets, and voids) are about 0.6\,\%, 7.6\,\%, 27.8\,\%, and 64.0\,\% respectively (i.e. with a void VFF $> 60$\%). We then classify the DT voids according to the position of their centres (see Fig.~\ref{fig:cosweb}).
For $R_V \geq 15\,h^{-1}$Mpc, over 80\,\% of the DT voids are in the void-like type, and more than 95\,\% of the DT voids are in either void-like or sheet-like structures, where the space is expanding on at least two directions, thus further confirming that the \textit{voids-in-voids} class of DT voids is following the emptiness.

As the void radius decreases, the DT void fraction in sheets, filaments, and knots display maxima at different sizes, and the highest occupation structure varies from voids to sheets, and then to filaments. The evolution of the DT void fraction indicates that the characteristic scale of the four cosmic web structures (voids, sheets, filaments, and knots) decreases in respective order. In particular, the void population in filamentary structures shows a maximum at $R_V \sim 3.5\,h^{-1}\mathrm{Mpc}$, which is consistent with the typical separation of galaxies along filaments ($\sim 7\,h^{-1}\mathrm{Mpc}$) obtained from the correlation function of galaxies in filaments \citep[][]{Tempel2014}. The position of the peaks in Fig.~\ref{fig:cosweb} reveals the typical separation of haloes in different types of cosmic web structures, since small DT voids connect four haloes with a maximum separation of $2 R_V$.

\section{Void Clustering}
\label{sec:clustering}

Due to the large volume of voids, they are typically regarded as rare objects in the universe. Therefore, the clustering properties of DT voids are seldom studied \citep[cf. however][]{Chan2014, Clampitt2015}.
However, the large number of DT voids permits us to have reliable clustering statistics, such as power spectra, or correlation functions. We present here some results on power spectra.

\subsection{Auto and cross power spectra}
\label{sec:pk}

As we have presented in previous sections, there are two completely different populations of DT voids, i.e., \textit{voids-in-voids} and \textit{voids-in-clouds} respectively. Therefore, instead of considering the whole void sample, let us consider the two populations separately.
Following \S~\ref{sec:numdens_nmean}, we define \textit{large} voids as $R_V \geq 16\,h^{-1}$Mpc voids. Similarly, we also define \textit{small} voids as $R_V < 8\,h^{-1}$Mpc, following the local densities shown in Fig.~\ref{fig:radens} and the cosmic web types in Fig.~\ref{fig:cosweb}.

\begin{figure}
\centering
\includegraphics[width=.47\textwidth]{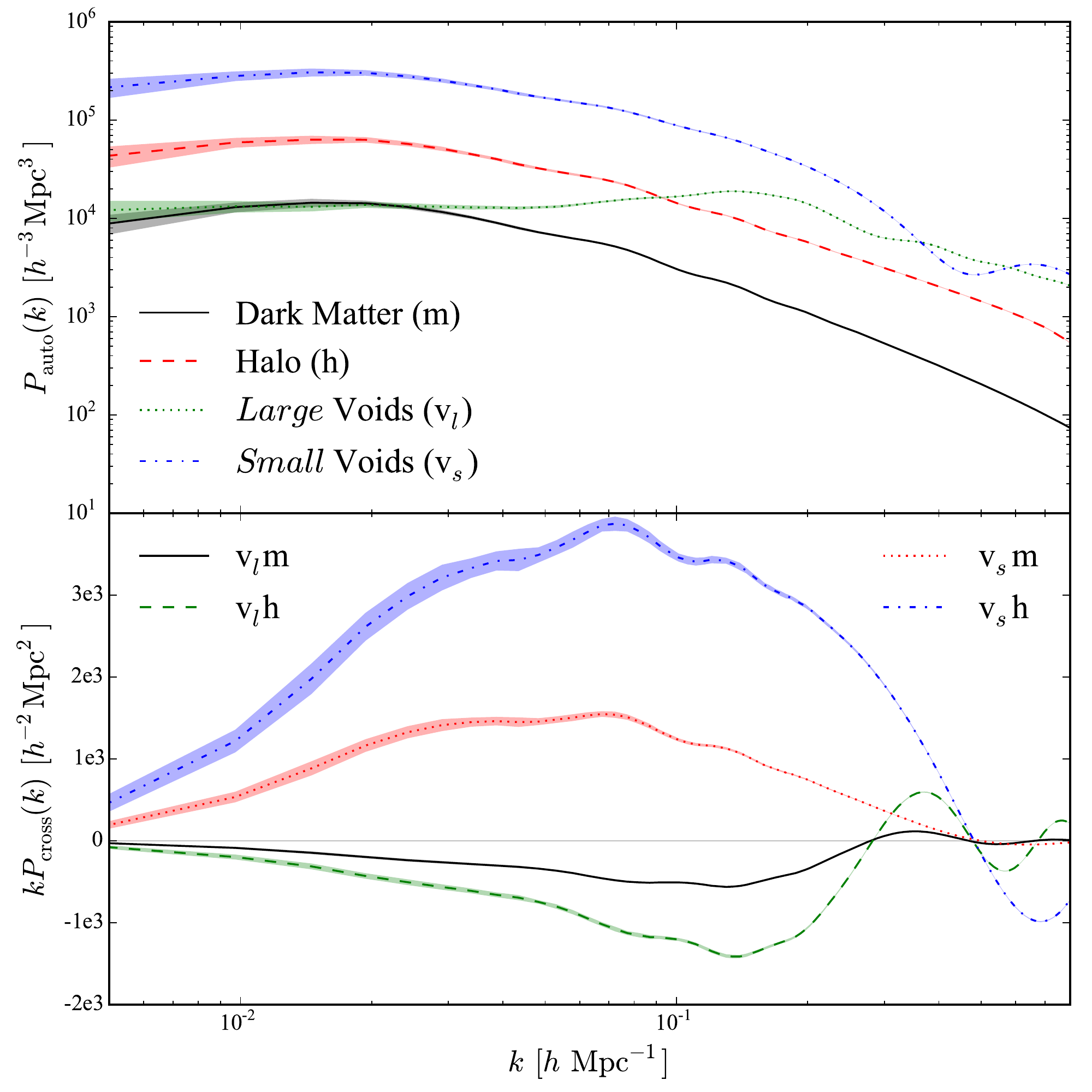}
\caption{Auto and cross power spectra for different tracers based on the \textsc{patchy} mocks. The shaded regions show the 1-$\sigma$ errors from 100 mocks.}
\label{fig:pk}
\end{figure}

We present the auto and cross power spectra for different tracers (dark matter, haloes, \textit{large} DT voids, and \textit{small} DT voids) based on 100 \textsc{patchy} mocks in Fig.~\ref{fig:pk}, where we adopt the Cloud-In-Cell (CIC) particle mass assignment scheme with a grid size of $960^3$ to perform fast Fourier Transform, and take into account the aliasing and shot noise corrections \citep[cf.][]{Jing2005}.

The two classes (\textit{small} and \textit{large} voids) show distinct power spectra, reflecting their different biases. The \textit{large} voids show a relatively low bias, and the amplitude of the auto power spectrum at small $k$ is very close to that of dark matter. On the contrary, the bias of the \textit{small} voids is fairly large, and the auto power spectrum of this class of voids follows the halo power spectrum quite well up to $k \sim 0.1\,h\,\mathrm{Mpc}^{-1}$. This is because \textsc{dive} selects groups of four haloes with a maximum separation, and works like the first phase of the cluster finding scheme developed by \citet[][]{Marinoni2002}. In this case, we are likely to find gravitational bound systems of haloes, which should have a higher bias than the overall population of haloes.
%The performance and implications of the DT algorithm as a group finder will be investigated in a future work.

Moreover, there are dips and oscillations at large $k$ ($k>0.1\,h\,\mathrm{Mpc}^{-1}$) for the auto and power spectra respectively, indicating patterns similar to the halo exclusion effect \citep[][]{Baldauf2013} and void exclusion effect \citep[][]{Chan2014}.
Therefore, even though the overlapping fraction is fairly high for the DT voids, there are still strong exclusion effects. This effect can be recognised also from the density profile shown in \S~\ref{sec:dens_pro}, where the profile is steep inside the voids, and implies a clear exclusion from other voids.

%\subsection{Bias and density profile revisit}
\subsection{Void bias and density profile}
\label{sec:dp_more}

\begin{figure*}
\begin{tabular}{cc}
\includegraphics[width=.47\textwidth]{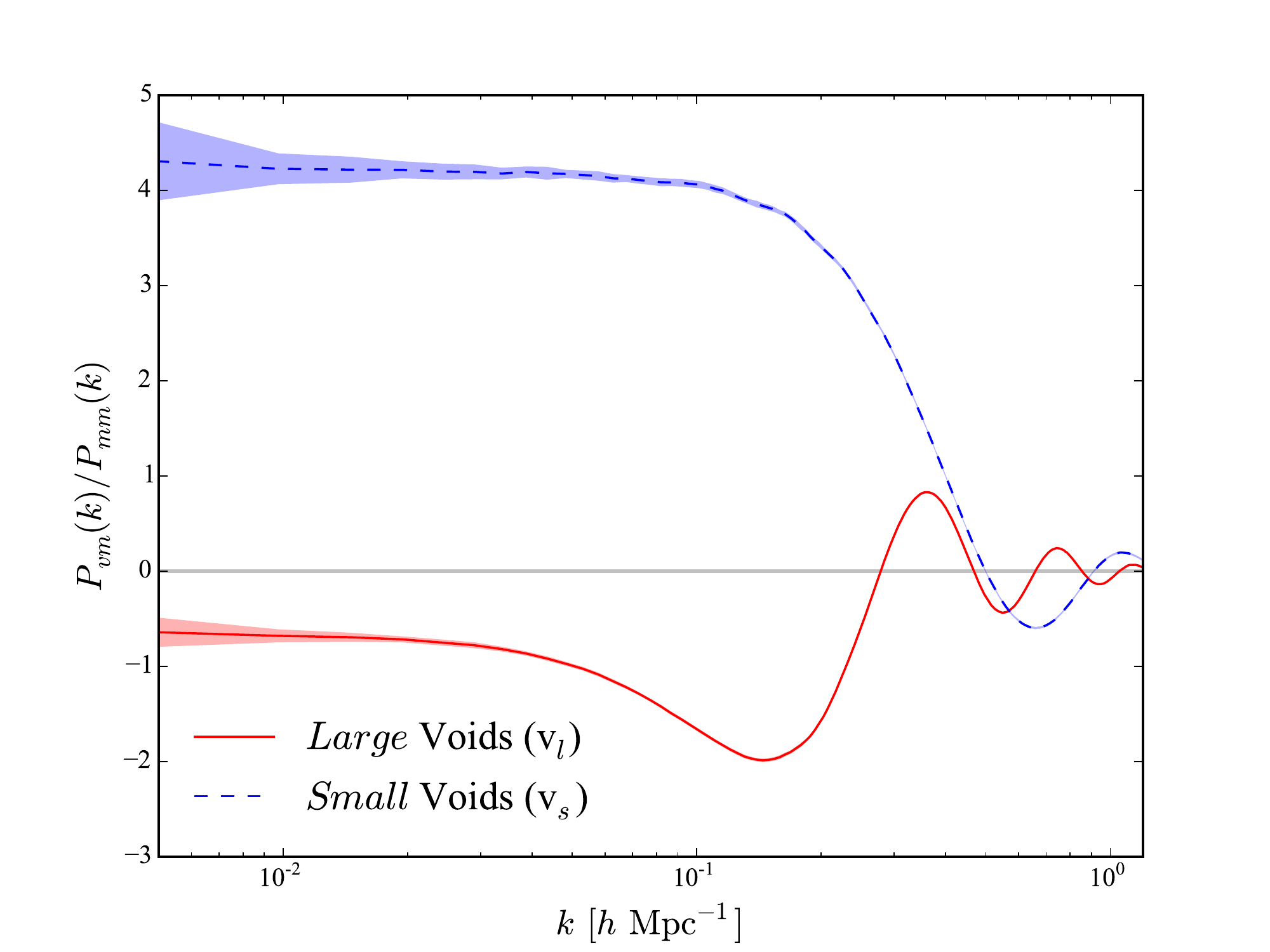}
\includegraphics[width=.47\textwidth]{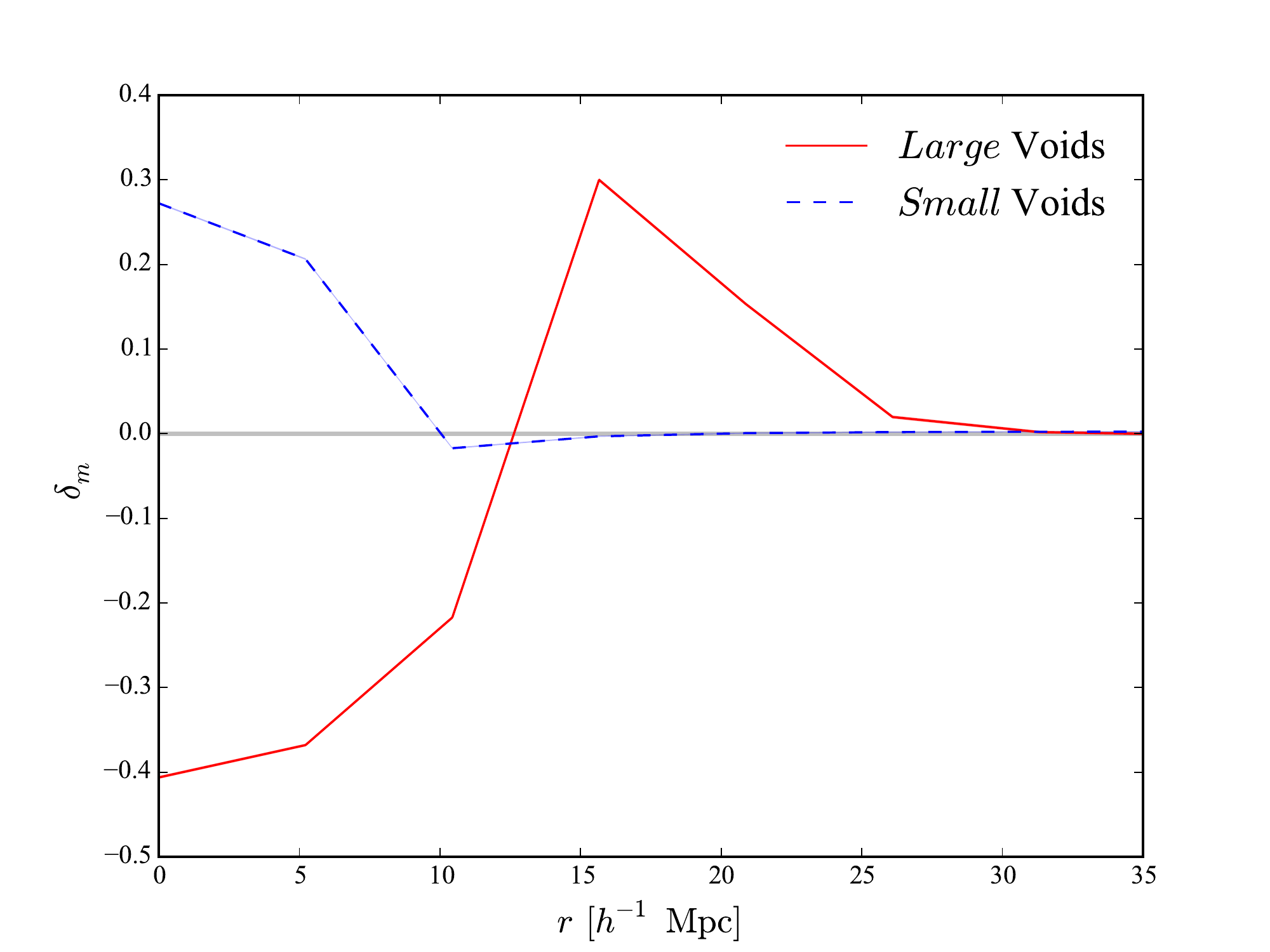}
\end{tabular}
\caption{Bias (\textit{left}) and radial density profile (\textit{right}) of DT voids, obtained from the auto and cross power spectra. The shaded regions show the 1-$\sigma$ errors from 100 mocks.}
\label{fig:bias_dp}
\end{figure*}

The clustering statistics permits us to have a quantitative measurement of the bias. Since voids suffer less from late-time nonlinear effects \citep[][]{vande1993}, we adopt the linear estimator of the scale-dependent bias of voids with respect to dark matter ($b_m (k)$):
\begin{equation}
b_m (k) = \frac{P_{vm}(k)}{P_{mm}(k)},
\end{equation}
where $P_{vm}(k)$ indicates the cross power spectrum between voids and dark matter, and $P_{mm}(k)$ is the auto power spectrum of dark matter.

We present the bias for the two classes of DT voids (\textit{large} and \textit{small} voids) on the \textit{left} panel of Fig.~\ref{fig:bias_dp}. In the linear regime (small $k$), the bias is nearly constant, but the two classes clearly show two distinct biases. The \textit{large} voids have a negative dark matter bias of about $-0.7$, indicating a very low density environment, while the bias of the \textit{small} voids is around 4.2, about twice the bias of LRGs \citep[][]{Tegmark2006}.

The radial density profile of the stacked voids describes the cross correlation between dark matter and void centres, thus the cross bias between dark matter and voids is essentially the Fourier counterpart of the density profile neglecting the shot noise \citep[][]{Chan2014, Hamaus2014b}:
\begin{equation}
u_v (k) = \frac{P_{v m}(k)}{P_{m m}(k)} \times \left| \frac{P_{m m}(k)}{P_{v m}(k)} \right|_{k \to 0} ,
\end{equation}
where $u_v (k)$ is the void density profile in Fourier space.
This equation is valid, since our dark matter samples is large enough ($\bar{n}_m^{-1} \ll P_{m m}(k)$).
We are then able to reproduce the density profile of the two classes of voids using the Fourier Transform of the scale-dependent bias.

The \textit{right} panel of Fig.~\ref{fig:bias_dp} shows the density profile obtained from the cross power spectra of 100 \textsc{patchy} mocks. The \textit{large} voids show a typical void density profile, while the \textit{small} void class show a very different one, in which the density contrast in the central part is very high. This confirms that the small radius voids found by \textsc{dive} are essentially groups of haloes.

%\subsubsection{Baryonic Acoustic Oscillations}
%\label{sec:pk_bao}

%To extract the BAO information from the power spectra, we compute the auto and cross power spectra for the wiggle free (non-wiggle) realisations of \textsc{patchy} mocks, and divide the power spectra of the with-wiggle realisations by the non-wiggle ones ($P(k) / P^{\rm nw} (k)$). The results are shown in Fig.~\ref{fig:pk_nw}. There are clear BAO signals in all three groups of DT voids. In general, the \textit{small} void group has the highest quality. The whole sample suffers from the void exclusion effect at large $k$, while the \textit{large} void group has fairly large errors at small $k$.

%Fitting and discussions ... (TBE.)

%\begin{figure*}
%\centering
%\includegraphics[width=.9\textwidth]{pk_nw}
%\caption{The ratio of the auto and cross power spectra of the \textsc{patchy} between the realisations with wiggles and without wiggles. The envelopes show the errors from 100 mocks.}
%\label{fig:pk_nw}
%\end{figure*}

%\subsubsection{Clustering of disjoint voids}
%\label{sec:clu_dj}

%TBE.

\section{Conclusions}
\label{sec:con}

We have presented a new parameter-free cosmological void finder based on Delaunay Triangulation, which emerges as the natural approach to be applied to discrete samples of haloes or galaxies. 
This method is very efficient on large data sets and we plan to use it for upcoming galaxy redshift surveys. In particular, the speed of the Delaunay Triangulation algorithm allows us to find the voids from catalogues with millions of objects covering huge volumes within minutes.

The complete set of Delaunay Triangulation voids (DT voids) shows high overlapping fractions, and cannot be considered as individual spherical under-dense regions in the universe. Interestingly their distribution depicts the aspherical shape of cosmic voids serving as proxies for the emptiness in the Universe.
With this void definition, the number of voids is about 7 times the halo population with a number density of $3.5\times 10^{-4}\,h^3\mathrm{Mpc}^{-3}$, which permits us to perform reliable statistical studies, in particular clustering analysis, based on voids.
Besides, the associated tetrahedra of DT voids constraint by Delaunay Triangulation involve high order correlation of the tracers, which allows us to extract additional information from distribution of galaxies when computing the correlation function of voids (cf. Kitaura et al., Liang et al.). The cosmological gain of DT voids will be discussed in a future paper (Chuang et al., in preparation).

We further studied the DT voids in a hierarchical perspective, where we interpret the maximum non-overlapping spheres as parent (disjoint) voids, and the overlapping ones as sub-voids, showing that the \textit{disjoint voids} alone are not able to represent all the under-dense regions of the universe, due to their low volume filling fraction. Moreover, we find that the larger parent voids have more and larger sub-voids.

Furthermore, we show that there exist two distinct populations of DT voids, i.e., \textit{voids-in-voids} and \textit{voids-in-clouds}, which can be accurately separated from each other based on the radius of the spheres.
Some of the properties we have found for the two void populations are:
\begin{enumerate}
\item
The volume filling fractions and number functions have different responses to the halo Redshift Space Distortion effects.
\item
The number function of DT voids is strongly correlated with the number density of tracers. However, small voids follow the population of tracers, while the number of large voids is anti-correlated with the tracer number density. Moreover, the halo and the large void number density fields are complementary to each other across cosmic volumes.
\item
The dark matter environment of the two species of voids are different. The centres of large voids are mostly in under-dense regions, which is consistent with the voids defined from the dark matter density field in a dynamical way. While the centres of small voids are in very dense regions, and only a small fraction is defined as voids using the dynamical method. Similarly, the density profile of large voids are deeper in the centre, while the small voids have relatively high centre density, as well as a high wall near the boundary.
\item
We find from clustering statistics that the two void populations have different bias. Large voids have negative bias, and small voids have high positive bias.
\end{enumerate}

We confirmed that the large DT voids (\textit{voids-in-voids}) follow the under-dense regions of the universe, while the small voids (\textit{voids-in-clouds}) are essentially groups of the tracers, which are gravitational bound systems efficiently found by \textsc{dive}.
We find also an universal intrinsic fluctuation of the void density profile, which may be related to the density fluctuation of the underlying dark matter field.

We apply this technique for further studies of voids in the large scale structure, to constrain cosmological parameters and alternative models of gravity. The BAO analyses of the two point correlation functions are detailed in accompanying papers (Kitaura et al., Liang et al., Chuang et al.), and the work on the power spectra is in preparation. Furthermore, \citet[][]{Hamaus2014b} present a novel scale found by void-galaxy cross correlation, and we are studying this characteristic scale using DT voids in the BOSS galaxy data.

\section*{acknowledgments}

CZ thanks He Wang and Qing Shu for some instructive discussions on the Delaunay Triangulation. CZ, CT, and YL acknowledge support from Tsinghua University, and 985 program. FSK thanks support from the Leibniz Society for the Karl-Schwarzschild fellowship.

We acknowledge the EREBOS/THEIA/GERAS clusters at AIP and the HYDRA HPC-cluster of IFT-UAM/CSIC for the computations. Fig.~\ref{fig:visual_all} and Fig.~\ref{fig:visual_dj} are made by the \textsc{Wolfram Mathematica}\footnote{\url{http://www.wolfram.com/mathematica/}} software.

% Create the reference section using BibTeX:
\bibliographystyle{mnras}
\bibliography{DTVoid}

\begin{thebibliography}{}
\makeatletter
\relax
\def\mn@urlcharsother{\let\do\@makeother \do\$\do\&\do\#\do\^\do\_\do\%\do\~}
\def\mn@doi{\begingroup\mn@urlcharsother \@ifnextchar [ {\mn@doi@}
  {\mn@doi@[]}}
\def\mn@doi@[#1]#2{\def\@tempa{#1}\ifx\@tempa\@empty \href
  {http://dx.doi.org/#2} {doi:#2}\else \href {http://dx.doi.org/#2} {#1}\fi
  \endgroup}
\def\mn@eprint#1#2{\mn@eprint@#1:#2::\@nil}
\def\mn@eprint@arXiv#1{\href {http://arxiv.org/abs/#1} {{\tt arXiv:#1}}}
\def\mn@eprint@dblp#1{\href {http://dblp.uni-trier.de/rec/bibtex/#1.xml}
  {dblp:#1}}
\def\mn@eprint@#1:#2:#3:#4\@nil{\def\@tempa {#1}\def\@tempb {#2}\def\@tempc
  {#3}\ifx \@tempc \@empty \let \@tempc \@tempb \let \@tempb \@tempa \fi \ifx
  \@tempb \@empty \def\@tempb {arXiv}\fi \@ifundefined
  {mn@eprint@\@tempb}{\@tempb:\@tempc}{\expandafter \expandafter \csname
  mn@eprint@\@tempb\endcsname \expandafter{\@tempc}}}

\bibitem[\protect\citeauthoryear{{Abel}, {Hahn}  \& {Kaehler}}{{Abel}
  et~al.}{2012}]{Abel2012}
{Abel} T.,  {Hahn} O.,   {Kaehler} R.,  2012, \mn@doi [\mnras]
  {10.1111/j.1365-2966.2012.21754.x}, \href
  {http://adsabs.harvard.edu/abs/2012MNRAS.427...61A} {427, 61}

\bibitem[\protect\citeauthoryear{{Aikio} \& {M{\"a}h{\"o}nen}}{{Aikio} \&
  {M{\"a}h{\"o}nen}}{1998}]{Aikio1998}
{Aikio} J.,  {M{\"a}h{\"o}nen} P.,  1998, \mn@doi [\apj] {10.1086/305509},
  \href {http://adsabs.harvard.edu/abs/1998ApJ...497..534A} {497, 534}

\bibitem[\protect\citeauthoryear{{Arag{\'o}n-Calvo}, {Jones}, {van de Weygaert}
   \& {van der Hulst}}{{Arag{\'o}n-Calvo} et~al.}{2007}]{Aragon2007}
{Arag{\'o}n-Calvo} M.~A.,  {Jones} B.~J.~T.,  {van de Weygaert} R.,   {van der
  Hulst} J.~M.,  2007, \mn@doi [\aap] {10.1051/0004-6361:20077880}, \href
  {http://adsabs.harvard.edu/abs/2007A%26A...474..315A} {474, 315}

\bibitem[\protect\citeauthoryear{{Baldauf}, {Seljak}, {Smith}, {Hamaus}  \&
  {Desjacques}}{{Baldauf} et~al.}{2013}]{Baldauf2013}
{Baldauf} T.,  {Seljak} U.,  {Smith} R.~E.,  {Hamaus} N.,   {Desjacques} V.,
  2013, \mn@doi [\prd] {10.1103/PhysRevD.88.083507}, \href
  {http://adsabs.harvard.edu/abs/2013PhRvD..88h3507B} {88, 083507}

\bibitem[\protect\citeauthoryear{{Berg{\'e}}, {Price}, {Amara}  \&
  {Rhodes}}{{Berg{\'e}} et~al.}{2012}]{Berge2012}
{Berg{\'e}} J.,  {Price} S.,  {Amara} A.,   {Rhodes} J.,  2012, \mn@doi
  [\mnras] {10.1111/j.1365-2966.2011.19888.x}, \href
  {http://adsabs.harvard.edu/abs/2012MNRAS.419.2356B} {419, 2356}

\bibitem[\protect\citeauthoryear{{Bernardeau}}{{Bernardeau}}{1996}]{Bernardeau1996}
{Bernardeau} F.,  1996, in {Coles} P.,  {Martinez} V.,   {Pons-Borderia} M.-J.,
   eds,  Astronomical Society of the Pacific Conference Series Vol. 94,
  Mapping, Measuring, and Modelling the Universe. p.~253 (\mn@eprint {}
  {astro-ph/9601083})

\bibitem[\protect\citeauthoryear{{Betancort-Rijo}, {Patiri}, {Prada}  \&
  {Romano}}{{Betancort-Rijo} et~al.}{2009}]{Betancort2009}
{Betancort-Rijo} J.,  {Patiri} S.~G.,  {Prada} F.,   {Romano} A.~E.,  2009,
  \mn@doi [\mnras] {10.1111/j.1365-2966.2009.15567.x}, \href
  {http://adsabs.harvard.edu/abs/2009MNRAS.400.1835B} {400, 1835}

\bibitem[\protect\citeauthoryear{{Braun}, {Thieulot}, {Fullsack}, {DeKool},
  {Beaumont}  \& {Huismans}}{{Braun} et~al.}{2008}]{Braun2008}
{Braun} J.,  {Thieulot} C.,  {Fullsack} P.,  {DeKool} M.,  {Beaumont} C.,
  {Huismans} R.,  2008, \mn@doi [Physics of the Earth and Planetary Interiors]
  {10.1016/j.pepi.2008.05.003}, \href
  {http://adsabs.harvard.edu/abs/2008PEPI..171...76B} {171, 76}

\bibitem[\protect\citeauthoryear{Br{\"o}nnimann, Fabri, Giezeman, Hert,
  Hoffmann, Kettner, Pion  \& Schirra}{Br{\"o}nnimann
  et~al.}{2015}]{Bronnimann2015}
Br{\"o}nnimann H.,  Fabri A.,  Giezeman G.-J.,  Hert S.,  Hoffmann M.,  Kettner
  L.,  Pion S.,   Schirra S.,  2015, in , {CGAL} User and Reference Manual,
  {4.6.3} edn, {CGAL Editorial Board}, \url
  {http://doc.cgal.org/4.6.3/Manual/packages.html#PkgKernel23Summary}

\bibitem[\protect\citeauthoryear{{Brunino}, {Trujillo}, {Pearce}  \&
  {Thomas}}{{Brunino} et~al.}{2007}]{Brunino2007}
{Brunino} R.,  {Trujillo} I.,  {Pearce} F.~R.,   {Thomas} P.~A.,  2007, \mn@doi
  [\mnras] {10.1111/j.1365-2966.2006.11282.x}, \href
  {http://adsabs.harvard.edu/abs/2007MNRAS.375..184B} {375, 184}

\bibitem[\protect\citeauthoryear{{Cai}, {Padilla}  \& {Li}}{{Cai}
  et~al.}{2015}]{Cai2015}
{Cai} Y.-C.,  {Padilla} N.,   {Li} B.,  2015, \mn@doi [\mnras]
  {10.1093/mnras/stv777}, \href
  {http://adsabs.harvard.edu/abs/2015MNRAS.451.1036C} {451, 1036}

\bibitem[\protect\citeauthoryear{{Cardiel}, {Jim{\'e}nez-Esteban}, {Alacid},
  {Solano}  \& {Aberasturi}}{{Cardiel} et~al.}{2011}]{Cardiel2011}
{Cardiel} N.,  {Jim{\'e}nez-Esteban} F.~M.,  {Alacid} J.~M.,  {Solano} E.,
  {Aberasturi} M.,  2011, \mn@doi [\mnras] {10.1111/j.1365-2966.2011.19464.x},
  \href {http://adsabs.harvard.edu/abs/2011MNRAS.417.3061C} {417, 3061}

\bibitem[\protect\citeauthoryear{{Cautun}, {van de Weygaert}  \&
  {Jones}}{{Cautun} et~al.}{2013}]{Cautun2013}
{Cautun} M.,  {van de Weygaert} R.,   {Jones} B.~J.~T.,  2013, \mn@doi [\mnras]
  {10.1093/mnras/sts416}, \href
  {http://adsabs.harvard.edu/abs/2013MNRAS.429.1286C} {429, 1286}

\bibitem[\protect\citeauthoryear{{Cautun}, {van de Weygaert}, {Jones}  \&
  {Frenk}}{{Cautun} et~al.}{2015}]{Cautun2015}
{Cautun} M.,  {van de Weygaert} R.,  {Jones} B.~J.~T.,   {Frenk} C.~S.,  2015,
  preprint (\mn@eprint {arXiv} {1501.01306})

\bibitem[\protect\citeauthoryear{{Cedr{\'e}s}, {Cepa}, {Bongiovanni},
  {Casta{\~n}eda}, {S{\'a}nchez-Portal}  \& {Tomita}}{{Cedr{\'e}s}
  et~al.}{2012}]{Cedres2012}
{Cedr{\'e}s} B.,  {Cepa} J.,  {Bongiovanni} {\'A}.,  {Casta{\~n}eda} H.,
  {S{\'a}nchez-Portal} M.,   {Tomita} A.,  2012, \mn@doi [\aap]
  {10.1051/0004-6361/201219571}, \href
  {http://adsabs.harvard.edu/abs/2012A%26A...545A..43C} {545, A43}

\bibitem[\protect\citeauthoryear{{Chan}, {Hamaus}  \& {Desjacques}}{{Chan}
  et~al.}{2014}]{Chan2014}
{Chan} K.~C.,  {Hamaus} N.,   {Desjacques} V.,  2014, \mn@doi [\prd]
  {10.1103/PhysRevD.90.103521}, \href
  {http://adsabs.harvard.edu/abs/2014PhRvD..90j3521C} {90, 103521}

\bibitem[\protect\citeauthoryear{{Chuang} et~al.,}{{Chuang}
  et~al.}{2015}]{Chuang2015}
{Chuang} C.-H.,  et~al., 2015, \mn@doi [\mnras] {10.1093/mnras/stv1289}, \href
  {http://adsabs.harvard.edu/abs/2015MNRAS.452..686C} {452, 686}

\bibitem[\protect\citeauthoryear{{Clampitt}, {Jain}  \&
  {S{\'a}nchez}}{{Clampitt} et~al.}{2015}]{Clampitt2015}
{Clampitt} J.,  {Jain} B.,   {S{\'a}nchez} C.,  2015, preprint (\mn@eprint
  {arXiv} {1507.08031})

\bibitem[\protect\citeauthoryear{{Colberg}, {Sheth}, {Diaferio}, {Gao}  \&
  {Yoshida}}{{Colberg} et~al.}{2005}]{Colberg2005}
{Colberg} J.~M.,  {Sheth} R.~K.,  {Diaferio} A.,  {Gao} L.,   {Yoshida} N.,
  2005, \mn@doi [\mnras] {10.1111/j.1365-2966.2005.09064.x}, \href
  {http://adsabs.harvard.edu/abs/2005MNRAS.360..216C} {360, 216}

\bibitem[\protect\citeauthoryear{{Dawson} et~al.,}{{Dawson}
  et~al.}{2013}]{Dawson2013}
{Dawson} K.~S.,  et~al., 2013, \mn@doi [\aj] {10.1088/0004-6256/145/1/10},
  \href {http://adsabs.harvard.edu/abs/2013AJ....145...10D} {145, 10}

\bibitem[\protect\citeauthoryear{Delaunay}{Delaunay}{1934}]{Delaunay1934}
Delaunay B.,  1934, Izv. Akad. Nauk SSSR, Otdelenie Matematicheskii i
  Estestvennyka Nauk, 7, 1

\bibitem[\protect\citeauthoryear{{Eisenstein} et~al.,}{{Eisenstein}
  et~al.}{2011}]{Eisenstein2011}
{Eisenstein} D.~J.,  et~al., 2011, \mn@doi [\aj] {10.1088/0004-6256/142/3/72},
  \href {http://adsabs.harvard.edu/abs/2011AJ....142...72E} {142, 72}

\bibitem[\protect\citeauthoryear{{El-Ad} \& {Piran}}{{El-Ad} \&
  {Piran}}{1997}]{Elad1997}
{El-Ad} H.,  {Piran} T.,  1997, \apj, \href
  {http://adsabs.harvard.edu/abs/1997ApJ...491..421E} {491, 421}

\bibitem[\protect\citeauthoryear{{Falck}, {Neyrinck}  \& {Szalay}}{{Falck}
  et~al.}{2012}]{Falck2012}
{Falck} B.~L.,  {Neyrinck} M.~C.,   {Szalay} A.~S.,  2012, \mn@doi [\apj]
  {10.1088/0004-637X/754/2/126}, \href
  {http://adsabs.harvard.edu/abs/2012ApJ...754..126F} {754, 126}

\bibitem[\protect\citeauthoryear{{Forero-Romero}, {Hoffman}, {Gottl{\"o}ber},
  {Klypin}  \& {Yepes}}{{Forero-Romero} et~al.}{2009}]{Forero2009}
{Forero-Romero} J.~E.,  {Hoffman} Y.,  {Gottl{\"o}ber} S.,  {Klypin} A.,
  {Yepes} G.,  2009, \mn@doi [\mnras] {10.1111/j.1365-2966.2009.14885.x}, \href
  {http://adsabs.harvard.edu/abs/2009MNRAS.396.1815F} {396, 1815}

\bibitem[\protect\citeauthoryear{{Foster} \& {Nelson}}{{Foster} \&
  {Nelson}}{2009}]{Foster2009}
{Foster} C.,  {Nelson} L.~A.,  2009, \mn@doi [\apj]
  {10.1088/0004-637X/699/2/1252}, \href
  {http://adsabs.harvard.edu/abs/2009ApJ...699.1252F} {699, 1252}

\bibitem[\protect\citeauthoryear{{Hahn}, {Porciani}, {Carollo}  \&
  {Dekel}}{{Hahn} et~al.}{2007}]{Hahn2007}
{Hahn} O.,  {Porciani} C.,  {Carollo} C.~M.,   {Dekel} A.,  2007, \mn@doi
  [\mnras] {10.1111/j.1365-2966.2006.11318.x}, \href
  {http://adsabs.harvard.edu/abs/2007MNRAS.375..489H} {375, 489}

\bibitem[\protect\citeauthoryear{{Hamaus}, {Wandelt}, {Sutter}, {Lavaux}  \&
  {Warren}}{{Hamaus} et~al.}{2014a}]{Hamaus2014b}
{Hamaus} N.,  {Wandelt} B.~D.,  {Sutter} P.~M.,  {Lavaux} G.,   {Warren} M.~S.,
   2014a, \mn@doi [Physical Review Letters] {10.1103/PhysRevLett.112.041304},
  \href {http://adsabs.harvard.edu/abs/2014PhRvL.112d1304H} {112, 041304}

\bibitem[\protect\citeauthoryear{{Hamaus}, {Sutter}  \& {Wandelt}}{{Hamaus}
  et~al.}{2014b}]{Hamaus2014a}
{Hamaus} N.,  {Sutter} P.~M.,   {Wandelt} B.~D.,  2014b, \mn@doi [Physical
  Review Letters] {10.1103/PhysRevLett.112.251302}, \href
  {http://adsabs.harvard.edu/abs/2014PhRvL.112y1302H} {112, 251302}

\bibitem[\protect\citeauthoryear{{Hoffman}, {Metuki}, {Yepes}, {Gottl{\"o}ber},
  {Forero-Romero}, {Libeskind}  \& {Knebe}}{{Hoffman}
  et~al.}{2012}]{Hoffman2012}
{Hoffman} Y.,  {Metuki} O.,  {Yepes} G.,  {Gottl{\"o}ber} S.,  {Forero-Romero}
  J.~E.,  {Libeskind} N.~I.,   {Knebe} A.,  2012, \mn@doi [\mnras]
  {10.1111/j.1365-2966.2012.21553.x}, \href
  {http://adsabs.harvard.edu/abs/2012MNRAS.425.2049H} {425, 2049}

\bibitem[\protect\citeauthoryear{{Hoyle} \& {Vogeley}}{{Hoyle} \&
  {Vogeley}}{2002}]{Hoyle2002}
{Hoyle} F.,  {Vogeley} M.~S.,  2002, \mn@doi [\apj] {10.1086/338340}, \href
  {http://adsabs.harvard.edu/abs/2002ApJ...566..641H} {566, 641}

\bibitem[\protect\citeauthoryear{{Icke} \& {van de Weygaert}}{{Icke} \& {van de
  Weygaert}}{1987}]{Icke1987}
{Icke} V.,  {van de Weygaert} R.,  1987, \aap, \href
  {http://adsabs.harvard.edu/abs/1987A%26A...184...16I} {184, 16}

\bibitem[\protect\citeauthoryear{Jamin, Pion  \& Teillaud}{Jamin
  et~al.}{2015}]{Jamin2015}
Jamin C.,  Pion S.,   Teillaud M.,  2015, in , {CGAL} User and Reference
  Manual, {4.6.3} edn, {CGAL Editorial Board}, \url
  {http://doc.cgal.org/4.6.3/Manual/packages.html#PkgTriangulation3Summary}

\bibitem[\protect\citeauthoryear{{Jennings}}{{Jennings}}{2012}]{Jennings2012}
{Jennings} E.,  2012, \mn@doi [\mnras] {10.1111/j.1745-3933.2012.01338.x},
  \href {http://adsabs.harvard.edu/abs/2012MNRAS.427L..25J} {427, L25}

\bibitem[\protect\citeauthoryear{{Jennings}, {Li}  \& {Hu}}{{Jennings}
  et~al.}{2013}]{Jennings2013}
{Jennings} E.,  {Li} Y.,   {Hu} W.,  2013, \mn@doi [\mnras]
  {10.1093/mnras/stt1169}, \href
  {http://adsabs.harvard.edu/abs/2013MNRAS.434.2167J} {434, 2167}

\bibitem[\protect\citeauthoryear{{Jing}}{{Jing}}{2005}]{Jing2005}
{Jing} Y.~P.,  2005, \mn@doi [\apj] {10.1086/427087}, \href
  {http://adsabs.harvard.edu/abs/2005ApJ...620..559J} {620, 559}

\bibitem[\protect\citeauthoryear{{Kitaura} \& {He{\ss}}}{{Kitaura} \&
  {He{\ss}}}{2013}]{Kitaura2013}
{Kitaura} F.-S.,  {He{\ss}} S.,  2013, \mn@doi [\mnras]
  {10.1093/mnrasl/slt101}, \href
  {http://adsabs.harvard.edu/abs/2013MNRAS.435L..78K} {435, L78}

\bibitem[\protect\citeauthoryear{{Kitaura}, {Yepes}  \& {Prada}}{{Kitaura}
  et~al.}{2014}]{Kitaura2014}
{Kitaura} F.-S.,  {Yepes} G.,   {Prada} F.,  2014, \mn@doi [\mnras]
  {10.1093/mnrasl/slt172}, \href
  {http://adsabs.harvard.edu/abs/2014MNRAS.439L..21K} {439, L21}

\bibitem[\protect\citeauthoryear{{Kitaura} et~al.,}{{Kitaura}
  et~al.}{2015a}]{Kitaura2015b}
{Kitaura} F.-S.,  et~al., 2015a, preprint (\mn@eprint {arXiv} {1509.06400})

\bibitem[\protect\citeauthoryear{{Kitaura}, {Gil-Mar{\'{\i}}n}, {Sc{\'o}ccola},
  {Chuang}, {M{\"u}ller}, {Yepes}  \& {Prada}}{{Kitaura}
  et~al.}{2015b}]{Kitaura2015}
{Kitaura} F.-S.,  {Gil-Mar{\'{\i}}n} H.,  {Sc{\'o}ccola} C.~G.,  {Chuang}
  C.-H.,  {M{\"u}ller} V.,  {Yepes} G.,   {Prada} F.,  2015b, \mn@doi [\mnras]
  {10.1093/mnras/stv645}, \href
  {http://adsabs.harvard.edu/abs/2015MNRAS.450.1836K} {450, 1836}

\bibitem[\protect\citeauthoryear{{Klypin}, {Yepes}, {Gottlober}, {Prada}  \&
  {Hess}}{{Klypin} et~al.}{2014}]{Klypin2014}
{Klypin} A.,  {Yepes} G.,  {Gottlober} S.,  {Prada} F.,   {Hess} S.,  2014,
  preprint (\mn@eprint {arXiv} {1411.4001})

\bibitem[\protect\citeauthoryear{{Lang}, {Holley-Bockelmann}  \&
  {Sinha}}{{Lang} et~al.}{2015}]{Lang2015}
{Lang} M.,  {Holley-Bockelmann} K.,   {Sinha} M.,  2015, \mn@doi [\apj]
  {10.1088/0004-637X/811/2/152}, \href
  {http://adsabs.harvard.edu/abs/2015ApJ...811..152L} {811, 152}

\bibitem[\protect\citeauthoryear{{Lee} \& {Park}}{{Lee} \&
  {Park}}{2009}]{Lee2009}
{Lee} J.,  {Park} D.,  2009, \mn@doi [\apjl] {10.1088/0004-637X/696/1/L10},
  \href {http://adsabs.harvard.edu/abs/2009ApJ...696L..10L} {696, L10}

\bibitem[\protect\citeauthoryear{{Marinoni}, {Davis}, {Newman}  \&
  {Coil}}{{Marinoni} et~al.}{2002}]{Marinoni2002}
{Marinoni} C.,  {Davis} M.,  {Newman} J.~A.,   {Coil} A.~L.,  2002, \mn@doi
  [\apj] {10.1086/343092}, \href
  {http://adsabs.harvard.edu/abs/2002ApJ...580..122M} {580, 122}

\bibitem[\protect\citeauthoryear{Medek, Bene{\v{s}}  \& Sochor}{Medek
  et~al.}{2007}]{Medek2007}
Medek P.,  Bene{\v{s}} P.,   Sochor J.,  2007, Journal of WSCG, 15, 107

\bibitem[\protect\citeauthoryear{{Nadathur}, {Hotchkiss}, {Diego}, {Iliev},
  {Gottl{\"o}ber}, {Watson}  \& {Yepes}}{{Nadathur}
  et~al.}{2015}]{Nadathur2015}
{Nadathur} S.,  {Hotchkiss} S.,  {Diego} J.~M.,  {Iliev} I.~T.,
  {Gottl{\"o}ber} S.,  {Watson} W.~A.,   {Yepes} G.,  2015, \mn@doi [\mnras]
  {10.1093/mnras/stv513}, \href
  {http://adsabs.harvard.edu/abs/2015MNRAS.449.3997N} {449, 3997}

\bibitem[\protect\citeauthoryear{{Neyrinck}}{{Neyrinck}}{2008}]{Neyrinck2008}
{Neyrinck} M.~C.,  2008, \mn@doi [\mnras] {10.1111/j.1365-2966.2008.13180.x},
  \href {http://adsabs.harvard.edu/abs/2008MNRAS.386.2101N} {386, 2101}

\bibitem[\protect\citeauthoryear{{Nuza}, {Kitaura}, {He{\ss}}, {Libeskind}  \&
  {M{\"u}ller}}{{Nuza} et~al.}{2014}]{Nuza2014}
{Nuza} S.~E.,  {Kitaura} F.-S.,  {He{\ss}} S.,  {Libeskind} N.~I.,
  {M{\"u}ller} V.,  2014, \mn@doi [\mnras] {10.1093/mnras/stu1746}, \href
  {http://adsabs.harvard.edu/abs/2014MNRAS.445..988N} {445, 988}

\bibitem[\protect\citeauthoryear{{Padilla}, {Ceccarelli}  \&
  {Lambas}}{{Padilla} et~al.}{2005}]{Padilla2005}
{Padilla} N.~D.,  {Ceccarelli} L.,   {Lambas} D.~G.,  2005, \mn@doi [\mnras]
  {10.1111/j.1365-2966.2005.09500.x}, \href
  {http://adsabs.harvard.edu/abs/2005MNRAS.363..977P} {363, 977}

\bibitem[\protect\citeauthoryear{{P{\'a}l} \& {Bakos}}{{P{\'a}l} \&
  {Bakos}}{2006}]{Pal2006}
{P{\'a}l} A.,  {Bakos} G.~{\'A}.,  2006, \mn@doi [\pasp] {10.1086/508573},
  \href {http://adsabs.harvard.edu/abs/2006PASP..118.1474P} {118, 1474}

\bibitem[\protect\citeauthoryear{{Pan}, {Vogeley}, {Hoyle}, {Choi}  \&
  {Park}}{{Pan} et~al.}{2012}]{Pan2012}
{Pan} D.~C.,  {Vogeley} M.~S.,  {Hoyle} F.,  {Choi} Y.-Y.,   {Park} C.,  2012,
  \mn@doi [\mnras] {10.1111/j.1365-2966.2011.20197.x}, \href
  {http://adsabs.harvard.edu/abs/2012MNRAS.421..926P} {421, 926}

\bibitem[\protect\citeauthoryear{{Patiri}, {Betancort-Rijo}, {Prada}, {Klypin}
  \& {Gottl{\"o}ber}}{{Patiri} et~al.}{2006}]{Patiri2006}
{Patiri} S.~G.,  {Betancort-Rijo} J.~E.,  {Prada} F.,  {Klypin} A.,
  {Gottl{\"o}ber} S.,  2006, \mn@doi [\mnras]
  {10.1111/j.1365-2966.2006.10305.x}, \href
  {http://adsabs.harvard.edu/abs/2006MNRAS.369..335P} {369, 335}

\bibitem[\protect\citeauthoryear{{Platen}, {van de Weygaert}  \&
  {Jones}}{{Platen} et~al.}{2007}]{Platen2007}
{Platen} E.,  {van de Weygaert} R.,   {Jones} B.~J.~T.,  2007, \mn@doi [\mnras]
  {10.1111/j.1365-2966.2007.12125.x}, \href
  {http://adsabs.harvard.edu/abs/2007MNRAS.380..551P} {380, 551}

\bibitem[\protect\citeauthoryear{{Platen}, {van de Weygaert}, {Jones}, {Vegter}
   \& {Calvo}}{{Platen} et~al.}{2011}]{Platen2011}
{Platen} E.,  {van de Weygaert} R.,  {Jones} B.~J.~T.,  {Vegter} G.,   {Calvo}
  M.~A.~A.,  2011, \mn@doi [\mnras] {10.1111/j.1365-2966.2011.18905.x}, \href
  {http://adsabs.harvard.edu/abs/2011MNRAS.416.2494P} {416, 2494}

\bibitem[\protect\citeauthoryear{{Ricciardelli}, {Quilis}  \&
  {Planelles}}{{Ricciardelli} et~al.}{2013}]{Ricciardelli2013}
{Ricciardelli} E.,  {Quilis} V.,   {Planelles} S.,  2013, \mn@doi [\mnras]
  {10.1093/mnras/stt1069}, \href
  {http://adsabs.harvard.edu/abs/2013MNRAS.434.1192R} {434, 1192}

\bibitem[\protect\citeauthoryear{{Rodr{\'{\i}}guez-Torres}
  et~al.,}{{Rodr{\'{\i}}guez-Torres} et~al.}{2015}]{Rodriguez2015}
{Rodr{\'{\i}}guez-Torres} S.~A.,  et~al., 2015, preprint (\mn@eprint {arXiv}
  {1509.06404})

\bibitem[\protect\citeauthoryear{{Romano-D{\'{\i}}az} \& {van de
  Weygaert}}{{Romano-D{\'{\i}}az} \& {van de Weygaert}}{2007}]{Romano2007}
{Romano-D{\'{\i}}az} E.,  {van de Weygaert} R.,  2007, \mn@doi [\mnras]
  {10.1111/j.1365-2966.2007.12190.x}, \href
  {http://adsabs.harvard.edu/abs/2007MNRAS.382....2R} {382, 2}

\bibitem[\protect\citeauthoryear{{Schaap} \& {van de Weygaert}}{{Schaap} \&
  {van de Weygaert}}{2000}]{Schaap2000}
{Schaap} W.~E.,  {van de Weygaert} R.,  2000, \aap, \href
  {http://adsabs.harvard.edu/abs/2000A%26A...363L..29S} {363, L29}

\bibitem[\protect\citeauthoryear{{Shandarin}, {Feldman}, {Heitmann}  \&
  {Habib}}{{Shandarin} et~al.}{2006}]{Shandarin2006}
{Shandarin} S.,  {Feldman} H.~A.,  {Heitmann} K.,   {Habib} S.,  2006, \mn@doi
  [\mnras] {10.1111/j.1365-2966.2006.10062.x}, \href
  {http://adsabs.harvard.edu/abs/2006MNRAS.367.1629S} {367, 1629}

\bibitem[\protect\citeauthoryear{{Shandarin}, {Habib}  \&
  {Heitmann}}{{Shandarin} et~al.}{2012}]{Shandarin2012}
{Shandarin} S.,  {Habib} S.,   {Heitmann} K.,  2012, \mn@doi [\prd]
  {10.1103/PhysRevD.85.083005}, \href
  {http://adsabs.harvard.edu/abs/2012PhRvD..85h3005S} {85, 083005}

\bibitem[\protect\citeauthoryear{{Sheth} \& {van de Weygaert}}{{Sheth} \& {van
  de Weygaert}}{2004}]{Sheth2004}
{Sheth} R.~K.,  {van de Weygaert} R.,  2004, \mn@doi [\mnras]
  {10.1111/j.1365-2966.2004.07661.x}, \href
  {http://adsabs.harvard.edu/abs/2004MNRAS.350..517S} {350, 517}

\bibitem[\protect\citeauthoryear{{Song} \& {Lee}}{{Song} \&
  {Lee}}{2009}]{Song2009}
{Song} H.,  {Lee} J.,  2009, \mn@doi [\apjl] {10.1088/0004-637X/701/1/L25},
  \href {http://adsabs.harvard.edu/abs/2009ApJ...701L..25S} {701, L25}

\bibitem[\protect\citeauthoryear{{Sousbie}}{{Sousbie}}{2011}]{Sousbie2011}
{Sousbie} T.,  2011, \mn@doi [\mnras] {10.1111/j.1365-2966.2011.18394.x}, \href
  {http://adsabs.harvard.edu/abs/2011MNRAS.414..350S} {414, 350}

\bibitem[\protect\citeauthoryear{{Springel}}{{Springel}}{2010}]{Springel2010}
{Springel} V.,  2010, \mn@doi [\mnras] {10.1111/j.1365-2966.2009.15715.x},
  \href {http://adsabs.harvard.edu/abs/2010MNRAS.401..791S} {401, 791}

\bibitem[\protect\citeauthoryear{{Sutter}, {Pisani}, {Wandelt}  \&
  {Weinberg}}{{Sutter} et~al.}{2014}]{Sutter2014}
{Sutter} P.~M.,  {Pisani} A.,  {Wandelt} B.~D.,   {Weinberg} D.~H.,  2014,
  \mn@doi [\mnras] {10.1093/mnras/stu1392}, \href
  {http://adsabs.harvard.edu/abs/2014MNRAS.443.2983S} {443, 2983}

\bibitem[\protect\citeauthoryear{{Tegmark} et~al.,}{{Tegmark}
  et~al.}{2006}]{Tegmark2006}
{Tegmark} M.,  et~al., 2006, \mn@doi [\prd] {10.1103/PhysRevD.74.123507}, \href
  {http://adsabs.harvard.edu/abs/2006PhRvD..74l3507T} {74, 123507}

\bibitem[\protect\citeauthoryear{{Tempel}, {Kipper}, {Saar}, {Bussov}, {Hektor}
   \& {Pelt}}{{Tempel} et~al.}{2014}]{Tempel2014}
{Tempel} E.,  {Kipper} R.,  {Saar} E.,  {Bussov} M.,  {Hektor} A.,   {Pelt} J.,
   2014, \mn@doi [\aap] {10.1051/0004-6361/201424418}, \href
  {http://adsabs.harvard.edu/abs/2014A%26A...572A...8T} {572, A8}

\bibitem[\protect\citeauthoryear{{Trujillo}, {Carretero}  \&
  {Patiri}}{{Trujillo} et~al.}{2006}]{Trujillo2006}
{Trujillo} I.,  {Carretero} C.,   {Patiri} S.~G.,  2006, \mn@doi [\apjl]
  {10.1086/503548}, \href {http://adsabs.harvard.edu/abs/2006ApJ...640L.111T}
  {640, L111}

\bibitem[\protect\citeauthoryear{{Zhao}, {Kitaura}, {Chuang}, {Prada}, {Yepes}
  \& {Tao}}{{Zhao} et~al.}{2015}]{Zhao2015}
{Zhao} C.,  {Kitaura} F.-S.,  {Chuang} C.-H.,  {Prada} F.,  {Yepes} G.,   {Tao}
  C.,  2015, \mn@doi [\mnras] {10.1093/mnras/stv1262}, \href
  {http://adsabs.harvard.edu/abs/2015MNRAS.451.4266Z} {451, 4266}

\bibitem[\protect\citeauthoryear{{van de Weygaert} \& {Schaap}}{{van de
  Weygaert} \& {Schaap}}{2007}]{vande2007}
{van de Weygaert} R.,  {Schaap} W.,  2007, preprint (\mn@eprint {arXiv}
  {0708.1441})

\bibitem[\protect\citeauthoryear{{van de Weygaert} \& {van Kampen}}{{van de
  Weygaert} \& {van Kampen}}{1993}]{vande1993}
{van de Weygaert} R.,  {van Kampen} E.,  1993, \mnras, \href
  {http://adsabs.harvard.edu/abs/1993MNRAS.263..481V} {263, 481}

\bibitem[\protect\citeauthoryear{{van de Weygaert} et~al.,}{{van de Weygaert}
  et~al.}{2011}]{vande2011}
{van de Weygaert} R.,  et~al., 2011, preprint (\mn@eprint {arXiv} {1110.5528})

\makeatother
\end{thebibliography}

\bsp
\label{lastpage}
\end{document}